\documentclass[%
 reprint,
superscriptaddress,
nofootinbib,
 amsmath,amssymb,
 aps,
prd,
]{revtex4-2}

\usepackage{graphicx}
\usepackage{dcolumn}
\usepackage{bm}

\usepackage{aas_macros}
\usepackage{xspace}
\usepackage{color}
\usepackage{hyperref}

\newcommand{\eg}{\mbox{e.\,g.,}\xspace}

\newcommand{\ie}{\mbox{i.\,e.,}\xspace}

\renewcommand{\vec}[1]{\bm{#1}}

\definecolor{darkgreen}{rgb}{0,0.5,0}
\definecolor{mypink1}{rgb}{0.858, 0.188, 0.478}

\begin{document}

\preprint{APS/123-QED}

\title{General framework for cosmological dark matter bounds using \(N\)-body simulations}

\author{Keir K. Rogers}
 \email{keir.rogers@fysik.su.se}

\affiliation{%
Oskar Klein Centre for Cosmoparticle Physics, Department of Physics, Stockholm University,\\
AlbaNova University Center, Stockholm 10691, Sweden}%

\author{Hiranya V. Peiris}
 \email{h.peiris@ucl.ac.uk}

\affiliation{
Department of Physics \& Astronomy, University College London, Gower Street, London WC1E 6BT, UK}%

\affiliation{%
Oskar Klein Centre for Cosmoparticle Physics, Department of Physics, Stockholm University,\\
AlbaNova University Center, Stockholm 10691, Sweden}

\date{\today}

\begin{abstract}
We present a general framework for obtaining robust bounds on the nature of dark matter using cosmological \(N\)-body simulations and Lyman-alpha forest data. We construct an emulator of hydrodynamical simulations, which is a flexible, accurate and computationally-efficient model for predicting the response of the Lyman-alpha forest flux power spectrum to different dark matter models, the state of the intergalactic medium (IGM) and the primordial power spectrum. The emulator combines a flexible parameterization for the small-scale suppression in the matter power spectrum arising in ``non-cold'' dark matter models, with an improved IGM model. We then demonstrate how to optimize the emulator for the case of ultra-light axion dark matter, presenting tests of convergence. We also carry out cross-validation tests of the accuracy of flux power spectrum prediction. This framework can be optimized for the analysis of many other dark matter candidates, \eg warm or interacting dark matter. Our work demonstrates that a combination of an optimized emulator and cosmological ``effective theories,'' where many models are described by a single set of equations, is a powerful approach for robust and computationally-efficient inference from the cosmic large-scale structure.
\end{abstract}

\maketitle

\section{\label{sec:intro}Introduction}

Determining the particle physics of the dark matter would not only explain an integral part of the standard cosmological model but can also point to new physics that will deepen and unify our understanding of nature. Following recent null results at high-energy colliders \citep{2018ARNPS..68..429B} and by direct detection \citep[\eg][]{2014arXiv1401.0216F}, it is timely to consider new theoretically well-motivated avenues that do not rely exclusively on the ``WIMP miracle'' \citep[weakly interacting massive particles;][]{1996PhR...267..195J}, as well as alternative experimental approaches beyond colliders and direct/indirect detection. Cosmological observations allow constraints on dark matter properties in physical regimes otherwise inaccessible in the laboratory \citep[\eg][]{2019BAAS...51c.134G, 2019arXiv190201055D}.

There exist many alternative proposals to the traditional WIMP, each with their own theoretical motivation. What unites a large number of these models are similar phenomenological features in cosmological structure formation (in particular, suppressed small-scale growth) constituting a deviation from the standard paradigm of cold, collisionless dark matter \citep{1982ApJ...263L...1P, blumenthal1984formation, davis1985evolution, 1986ApJ...304...15B}. These include, \eg light axions \citep{PhysRevLett.38.1440, PhysRevLett.40.223, PhysRevLett.40.279, PRESKILL1983127, ABBOTT1983133, DINE1983137, 2000PhRvL..85.1158H}; warm dark matter (particles with an associated free-streaming scale), \eg sterile neutrinos resonantly produced \citep{ENQVIST1990531} or as a decay product in the early Universe \citep[\eg][]{2016JCAP...11..038K}; mixed models, \eg of cold and warm dark matter \citep{2009JCAP...05..012B}; light (sub-GeV) particle candidates with non-negligible scattering with baryons \citep{2014PhRvD..89b3519D}, \eg those with electric and magnetic dipole moments \citep{2004PhRvD..70h3501S} or massive boson exchange \citep{2011PhRvL.106q1302L}; and the wide set of models described by the effective theory of structure formation \citep{2016PhRvD..93l3527C}, \eg with dark radiative components \citep{2009PhRvD..79b3519A} or self-interactions \citep{2000PhRvL..84.3760S}.

The Lyman-alpha forest (as a tracer of the small-scale, high-redshift linear matter power spectrum \citep{2019MNRAS.489.2247C}), often in combination with larger-scale probes like the cosmic microwave background (CMB) and baryon acoustic oscillations, provides competitive bounds on these models  \citep{2017PhRvD..96b3522I, 2019arXiv191209397G, 2020JCAP...04..038P, 2018PhRvD..98h3540M, 2009JCAP...05..012B, 2014PhRvD..89b3519D, 2018PhRvD..97j3530X, 2019JCAP...10..055A, 2017PhRvL.119c1302I, 2017MNRAS.471.4606A, 2017PhRvD..96l3514K}, to which we will refer in general as ``non-cold'' dark matter, following Ref.~\cite{2017JCAP...11..046M}. The challenge in getting robust bounds using the Lyman-alpha forest and other probes of the cosmic large-scale structure (\eg galaxies, weak lensing, 21 cm) is the computational cost of modeling the data with sufficient accuracy. This often requires \(N\)-body simulations \citep{1980MNRAS.192..321D, 1981MNRAS.194..503E}, sometimes with hydrodynamics \citep{2001NewA....6...79S} which adds even further computational cost. This makes ``brute-force'' sampling of the parameter space (as required by \eg Markov chain Monte Carlo methods; MCMC) computationally infeasible. Further, the many different dark matter and cosmological models that can be tested each require, in principle, a new set of simulations for each analysis, which can make inefficient use of computational resources. Much effort has gone into developing cosmological ``effective theories,'' where many different physical models are described by a single set of equations \citep[\eg][]{2016PhRvD..93l3527C, 2019MNRAS.488.2121C}. These are designed to address the issue of testing many physical theories in an efficient way.

In this work, we present a framework for cosmological dark matter bounds, specifically using the Lyman-alpha forest, which makes the testing of new models robust, accurate and computationally efficient. This combines a general model for the effect of non-cold dark matter (nCDM) on the matter power spectrum presented in Ref.~\cite{2017JCAP...11..046M} with the Bayesian emulator optimization we presented in Ref.~\cite{2019JCAP...02..031R, 2019JCAP...02..050B}. An ``emulator'' is a flexible model that can characterize the data accurately and is computationally cheap to evaluate \citep{rasmussen2003gaussian}. This means it can be easily called within MCMC, making parameter inference feasible. It is optimized using a small set of ``training'' simulations, the selection of which is critical (see \S~\ref{sec:method_ULA} for discussion on Bayesian emulator optimization). Emulators have found wide use in cosmological studies using the large-scale structure \citep{2019JCAP...02..031R, 2019JCAP...02..050B, Heitmann:2009, Kwan:2013, 2018arXiv180405867Z, Liu:2015, Petri:2015, 2018arXiv181109141J, 2018arXiv180405866M, 2018ApJ...852...22W, 2019MNRAS.490.4826G}.

We build an emulator for the Lyman-alpha forest flux power spectrum as a function of the nCDM model and other nuisance parameters. This initial base emulator can then be optimized for the analysis of specific dark matter models. Here, we consider the example of ultra-light axion (ULA) dark matter, but the procedure will be equivalent for any of the dark matter candidates which can be characterized by the nCDM model (see above). We optimize the construction of the emulator training set using Bayesian optimization, which iteratively adds training data in order to determine the peak of the posterior distribution accurately while exploring the parameter space fully \citep{10.1115/1.3653121}. This allows tests of the robustness of dark matter bounds with respect to the accuracy of the emulator model. Previous dark matter and cosmological bounds from the Lyman-alpha forest \citep[\eg][]{2017PhRvD..96b3522I, 2019arXiv191209397G, 2020JCAP...04..038P, 2009JCAP...05..012B, 2014PhRvD..89b3519D, 2018PhRvD..97j3530X, 2019JCAP...10..055A, 2017PhRvL.119c1302I, 2017MNRAS.471.4606A, 2017PhRvD..96l3514K, 2005PhRvD..71j3515S, 2015JCAP...02..045P, 2015JCAP...11..011P, 2017JCAP...06..047Y, 2019JCAP...07..017C} have relied on linear interpolation of simulated flux power spectra around a fiducial point. Ref.~\cite{2018PhRvD..98h3540M} interpolate using ``ordinary kriging'', which has some similarity to the method we consider here, although they do not model full parameter covariance nor test for convergence using Bayesian optimization (see \S~\ref{sec:method_ULA}). The emulator model we use makes fewer assumptions and is more robust in its statistical modeling, which can remove bias in power spectrum estimation and strengthen parameter constraints \citep{2019JCAP...02..050B}. Even for cosmological datasets where emulators are already used, the combination of convergence tests using Bayesian optimization and effective theories as described above can provide significant benefits.

In \S~\ref{sec:emu_nCDM}, we present the general emulator for nCDM models and then in \S~\ref{sec:emu_ULA}, we present the example of optimizing the emulator for the study of ULA dark matter. We conduct tests of emulator convergence and cross-validation in \S~\ref{sec:results}, discuss in \S~\ref{sec:discussion} and conclude in \S~\ref{sec:concs}.

\section{\label{sec:emu_nCDM}Non-cold dark matter emulator}

In this section, we present a general emulator of hydrodynamical simulations for non-cold dark matter models using the Lyman-alpha forest flux power spectrum as the constraining data. This initial emulator is not tied to a specific dark matter model, but rather forms the base for Bayesian optimization for a particular model (see \S~\ref{sec:emu_ULA} for the example of ultra-light axions). This framework ensures the computational efficiency, theoretical accuracy and maximum precision for dark matter bounds using the cosmic large-scale structure. In \S~\ref{sec:model_nCDM}, we explain the non-cold dark matter model we employ, first introduced by Ref.~\cite{2017JCAP...11..046M}. We present the details of our cosmological hydrodynamical simulations in \S~\ref{sec:sims} and the intergalactic medium and background cosmology model which must be marginalized in \S~\ref{sec:model_cosmo}. In \S~\ref{sec:method_nCDM}, we discuss the method of emulation by Gaussian processes and how this connects to Bayesian inference.

\subsection{\label{sec:model_nCDM}Non-cold dark matter model}

In order to build an emulator-inference framework which is as general and powerful as possible, we employ the non-cold dark matter (nCDM) model first introduced by Ref.~\cite{2017JCAP...11..046M}. The model is defined by a ``transfer function'',
\begin{equation}\label{eq:transfer}
\begin{split}
T(k) &\equiv \sqrt{\frac{P_\mathrm{nCDM}(k)}{P_\mathrm{CDM}(k)}}\\
&= [1 + (\alpha k)^\beta]^\gamma.
\end{split}
\end{equation}
\(P_\mathrm{CDM}(k)\) and \(P_\mathrm{nCDM}(k)\) are the linear matter power spectra respectively for CDM and nCDM models; \(k\) is the comoving wavenumber (length-scale) in \(h\,\mathrm{Mpc}^{-1}\); and \([\alpha, \beta, \gamma]\) are the free parameters of the model, where \(\alpha\) is in units of \(h^{-1}\,\mathrm{Mpc}\). In our analysis, we define \(T(k)\) at the initial conditions of our cosmological simulations (see \S~\ref{sec:sims}), \ie at \(z = 99\). It is a generalization of the fitting function for a thermal warm dark matter (WDM) transfer function \citep{2001ApJ...556...93B}, to which Eq.~\eqref{eq:transfer} reduces when \(\beta = 2\nu\), \(\gamma = -\frac{5}{\nu}\) and \(\nu = 1.12\). By having two further degrees of freedom, the general model in Eq.~\eqref{eq:transfer} can characterize the matter power spectra for many models beyond the thermal WDM case, \eg non-thermal WDM, like sterile neutrinos resonantly produced or from particle decays, mixed models of cold and warm dark matter, axion-like particles, light (sub-GeV) dark matter coupling to baryons, dark matter coupling to dark radiation or self-interacting dark matter \citep{ENQVIST1990531, 2016JCAP...11..038K, 2009JCAP...05..012B, 2014PhRvD..89b3519D, 2004PhRvD..70h3501S, 2011PhRvL.106q1302L, 2016PhRvD..93l3527C, 2009PhRvD..79b3519A, 2000PhRvL..84.3760S}. Many of these models are explored in Ref.~\cite{2017JCAP...11..046M}; we will consider the example of ultra-light axions in \S~\ref{sec:emu_ULA}.

\begin{figure}
\includegraphics[width=\columnwidth]{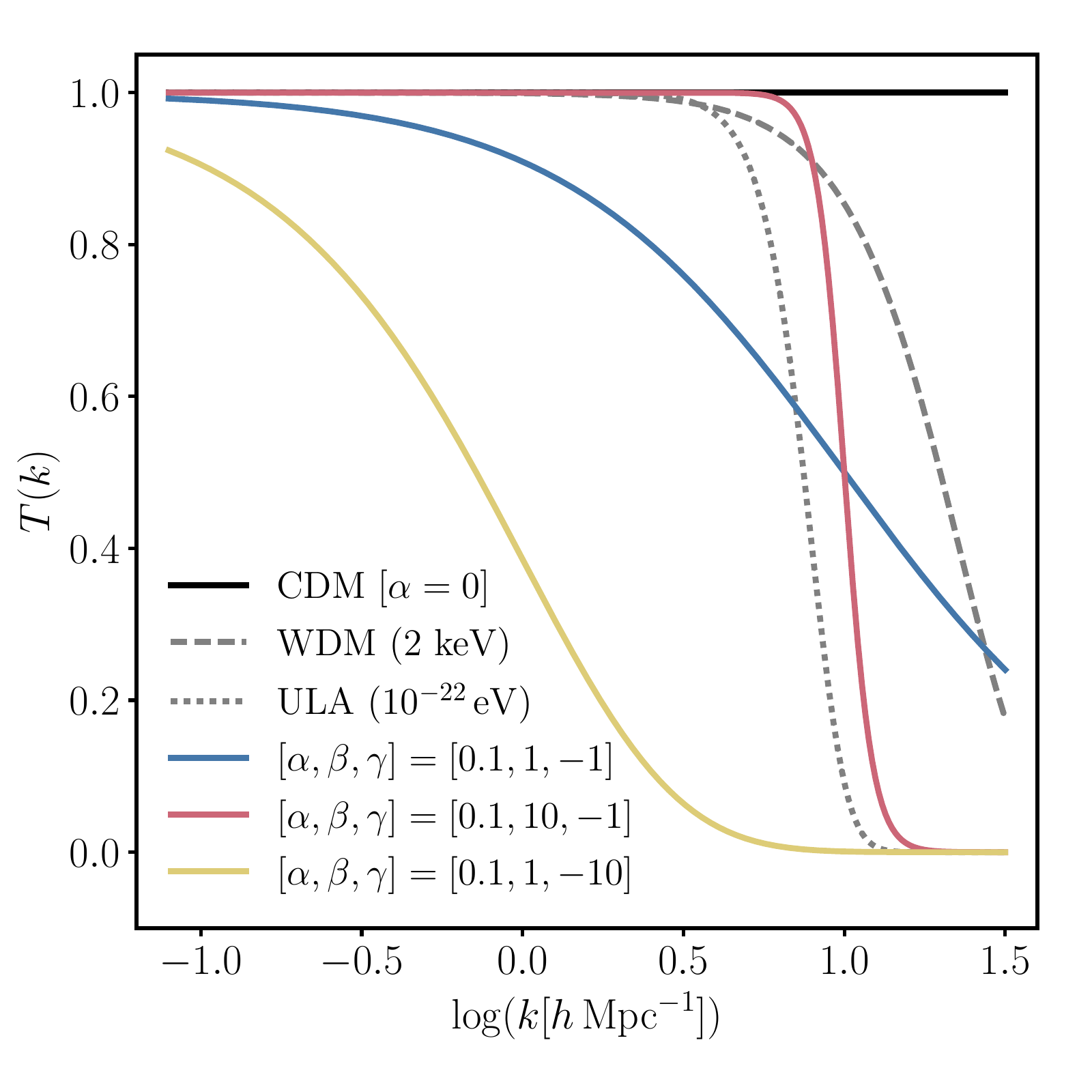}
\caption{\label{fig:transfer}Transfer functions \(T(k)\) (as defined in Eq.~\eqref{eq:transfer} at \(z = 99\)) for various configurations of the non-cold dark matter (nCDM) model as a function of comoving distance wavenumber \(k\). The colored lines show some extremal values of our prior range, while the black lines show nCDM fits for cold dark matter (\(\alpha = 0\)), 2 keV warm dark matter (\([\alpha, \beta, \gamma] = [0.02, 2.2, -4.5]\)) and \(10^{-22}\) eV ultra-light axion dark matter \([\alpha, \beta, \gamma] = [0.1, 5.2, -3.4]\).}
\end{figure}
We consider \(\alpha \in [0, 0.1]\), \(\beta \in [1, 10]\), \(\gamma \in [-10, 0]\). Fig.~\ref{fig:transfer} shows some example nCDM transfer functions. In particular, we show the cold dark matter limit with \(\alpha = 0\); nCDM model fits for the transfer function of a 2 keV WDM particle \([\alpha, \beta, \gamma] = [0.02, 2.2, -4.5]\) and a \(10^{-22}\) eV ULA particle \([\alpha, \beta, \gamma] = [0.1, 5.2, -3.4]\) (more details in \S~\ref{sec:emu_ULA}); and some extremal values of the nCDM parameters within our prior volume. In general, the nCDM model can characterize dark matter candidates which suppress the growth of structure on small scales and hence form a cut-off in the matter power spectrum. By spanning the three free parameters of the model, it is possible to characterize different scales and shapes of this cut-off, which can map to different particle models, as shown by the WDM and ULA examples in Fig.~\ref{fig:transfer}. \(\alpha\) is the main parameter determining the scale of the suppression and as it increases, the cut-off moves to larger scales (smaller \(k\)). \(\beta\) and \(\gamma\) are the main parameters determining the shape of the suppression. As \(\beta\) increases, the cut-off becomes sharper, in particular flatter before \(T^2 = 0.5\); as \(\gamma\) increases, the cut-off also becomes sharper, in particular flatter after \(T^2 = 0.5\).

\subsection{\label{sec:sims}Cosmological hydrodynamical simulations}

Our observable is the Lyman-alpha forest flux power spectrum at \(z = [4.2, 4.6, 5.0]\). In order to model this, we run cosmological hydrodynamical simulations of the intergalactic medium (IGM; from which the forest is sourced) using the publicly-available\footnote{\url{https://github.com/MP-Gadget/MP-Gadget}.} code \texttt{MP-Gadget} \citep{yu_feng_2018_1451799, 2001NewA....6...79S, 2005MNRAS.364.1105S}. We evolve \(512^3\) particles each of dark matter and gas in a \((10\,h^{-1}\,\mathrm{Mpc})^3\) box (see tests of convergence with respect to particle number and box size in Appendix \ref{sec:appendix_convergence}) from \(z = 99\) to \(z = 4.2\), saving snapshots of particle data in particular at \(z = [4.2, 4.6, 5.0]\). We include the effect of the nCDM models we study by generating the initial conditions of each simulation using a suppressed matter power spectrum as modified by the transfer function in Eq.~\eqref{eq:transfer}; we discuss the implications for the nCDM models we can characterize in \S~\ref{sec:discussion}. We focus computational resources on the low-density gas that forms the Lyman-alpha forest by enabling the \texttt{quick-lya} flag, which converts gas particles at overdensities \(> 1000\) and with temperatures \(< 10^5\) K into collisionless particles \citep{10.1111/j.1365-2966.2004.08224.x}. This yields significant computational efficiency gains with a negligible impact on the accuracy of Lyman-alpha forest statistics.

From each particle snapshot, we generate 32000 mock spectra (with pixel widths \(\Delta v = 1\,\mathrm{km}\,\mathrm{s}^{-1}\)) with only the Lyman-alpha absorption line and calculate the 1D flux power spectrum using \texttt{fake\_spectra} \citep{2017ascl.soft10012B}. The 1D Lyman-alpha forest flux power spectrum measures correlations in Lyman-alpha absorption lines along the line-of-sight only (\ie integrated over transverse directions). It is defined as \(P_\mathrm{f} (k_\mathrm{f}) \equiv \int \frac{\mathrm{d}\vec{k_\mathrm{f}^\perp}}{(2 \pi)^2}\,P_\mathrm{f}^\mathrm{3D} (\vec{k_\mathrm{f}})\). Here, \(k_\mathrm{f}\) and \(\vec{k_\mathrm{f}^\perp}\) are respectively the line-of-sight and transverse components of the full velocity wavevector \(\vec{k_\mathrm{f}}\) and have inverse-velocity units (\eg \(\mathrm{s}\,\mathrm{km}^{-1}\)); \(P_\mathrm{f}^\mathrm{3D} (\vec{k_\mathrm{f}})\) is the 3D flux power spectrum. 

We estimate \(P_\mathrm{f} (k_\mathrm{f})\) first in each mock spectrum using a fast Fourier transform (FFT): \(P_\mathrm{f}^i (k_\mathrm{f}) = |\hat{\delta}_\mathrm{f} (k_\mathrm{f})|^2\), where \(\hat{\delta}_\mathrm{f} (k_\mathrm{f})\) is the 1D FFT of \(\delta_\mathrm{f} (v) = \frac{\mathcal{F}}{\langle\mathcal{F}\rangle} - 1\). \(\mathcal{F} = e^{-\tau}\) is the fraction of emitted quasar flux transmitted, \(\langle\mathcal{F}\rangle\) is the mean flux across all spectra in a given box, \(\tau\) is the optical depth and the effective optical depth \(\tau_\mathrm{eff} = -\ln \langle\mathcal{F}\rangle\); \(v\) is the line-of-sight velocity. In Ref.~\cite{2019ApJ...872..101B}, the flux is normalized to a ``rolling mean flux'' (averaged at each pixel using a finite window). This attempts to correct for redshift evolution within individual spectra and mis-modeling of the quasar emission continuum; we test the impact on our results in Appendix \ref{sec:appendix_mean}. Finally, we then average the flux power spectra over all mock spectra and correct for the pixel window function \citep{2013A&A...559A..85P}: \(P_\mathrm{f} (k_\mathrm{f}) = \left\langle\frac{P_\mathrm{f}^i (k_\mathrm{f})}{W^2(k_\mathrm{f}, \Delta v)}\right\rangle_i\). The pixel window function is \(W(k_\mathrm{f}, \Delta v) = \frac{\sin(k_\mathrm{f} \Delta v / 2)}{k_\mathrm{f} \Delta v / 2}\); note that we do not simulate the finite resolution of the spectrograph as this does not commute with the rescaling of mean fluxes (see \S~\ref{sec:IGM_input}) and we compare only to data where the spectrographic resolution has been corrected (we test the impact of imperfect resolution modeling in Appendix \ref{sec:appendix_res}).

We calculate (and emulate; see \S~\ref{sec:method_nCDM}) the flux power spectrum at the (real-valued) discrete Fourier transform sample frequencies of the simulation box with wavenumbers in inverse-distance units (\(h\,\mathrm{Mpc}^{-1}\)). This is because the conversion to velocity units depends on redshift and the energy density by the Hubble law. Since the matter energy density will be varied in our base emulator (see \S~\ref{sec:cosmo}), this would mean that the velocity wavenumbers of each simulation would vary across the emulator and the modes would not be consistently distributed. In the likelihood function (see \S~\ref{sec:method_nCDM}), we linearly interpolate to velocity wavenumbers.

\subsection{\label{sec:model_cosmo}Intergalactic medium and cosmological model}

In order to study nCDM models using the flux power spectrum, we must marginalize over uncertainty in the thermal and ionization history of the IGM (which partly derives from the uncertain nature of cosmic reionization), as well as relevant aspects of the cosmological model.

\subsubsection{\label{sec:pressure_smoothing}Pressure smoothing}

The energy deposited in the IGM by the background of ionizing ultra-violet (UV) photons and the finite temperature of the IGM (\(\sim 10^4\,\mathrm{K}\) at \(z \sim 5\)) leads to a few different observable effects in the flux power spectrum. The heat in the IGM smoothes the spatial distribution of the gas below a characteristic scale, where the gas pressure balances gravitational clustering. A simple, classical interpretation defines a Jeans pressure smoothing scale, which is a function of the instantaneous gas temperature \cite[\eg][]{binney2011galactic}. However, in an expanding universe with an evolving thermal state, the relevant pressure smoothing scale depends on the integrated history of the gas heating as spatial fluctuations from earlier times expand or fail to collapse depending on the temperature at that epoch. This ``filtering length'' in general deviates from the Jeans length, being typically greater before reionization and less afterwards (\(\sim 100\) comoving kpc at \(z \sim 5\)).

In order to track the heat deposited in our simulations and the resulting filtering scale, we use the integrated energy deposited into the gas per unit mass at the mean density,
\begin{equation}\label{eq:u_0}
u_0 (z) = \int_{z_\mathrm{end}}^{z_\mathrm{ini}} \frac{\mathrm{d}z}{H(z) (1+z)}\,\frac{\sum_i n_i \epsilon_i}{\bar{\rho}},
\end{equation}
first introduced by Ref.~\cite{2016MNRAS.463.2335N}. Here, \(n_i\) are the number densities for each species \(i \in\) [HI, HeI, HeII], \(\bar{\rho}\) is the mean mass density, \(H(z)\) is the Hubble parameter and \([z_\mathrm{end}, z_\mathrm{ini}]\) sets the redshift range for integration. Ref.~\cite{2019ApJ...872..101B} determined redshift ranges for integration at \(z = [4.2, 4.6, 5.0]\), such that \(u_0\) most strongly correlates with the flux power spectrum. We use those redshift ranges: at \(z = 4.2\), \([z_\mathrm{end}, z_\mathrm{ini}] = [4.2, 12]\); at \(z = 4.6\), \([z_\mathrm{end}, z_\mathrm{ini}] = [4.6, 13]\); at \(z = 5.0\), \([z_\mathrm{end}, z_\mathrm{ini}] = [6, 13]\). In general, a delay between heat injection and a change in the gas distribution is expected, but this delay can be reduced in the flux power spectrum owing to non-linear redshift-space distortions arising from gas peculiar velocities; Ref.~\cite{2019ApJ...872..101B} posits this as an explanation for the different \(z_\mathrm{end}\). Otherwise, the values of \(z_\mathrm{ini}\) are consistent with the gas gradually losing sensitivity to heating at earlier times.

\subsubsection{\label{sec:TDR}Temperature-density relation}

The small-scale flux power spectrum is sensitive to the instantaneous thermal state of the IGM in particular owing to the thermal broadening of absorption lines. After the gas is heated by the (re-)ionization front, it cools (for \(\Delta z \sim 1 - 2\)) towards a thermal asymptote set largely by the balance between UV background (UVB) photo-heating and adiabatic cooling owing to cosmological expansion \citep[\eg][]{2009ApJ...694..842M}. Further, the overwhelming majority of the low-density gas (\(\Delta \lesssim 10\)) to which the forest is sensitive follows a tight power-law temperature-density relation \citep{1997MNRAS.292...27H}, where gas temperature
\begin{equation}\label{eq:TDR}
T(z) = T_0(z) \Delta^{\widetilde{\gamma}(z) - 1}.
\end{equation}
This relation has two free parameters: the temperature at mean density \(T_0 (z)\) and a slope \(\widetilde{\gamma}(z)\). The latter is generally expected to be \(\sim 1\) (in particular immediately after an ionization event) and to asymptote towards 1.6; however, \(\widetilde{\gamma}(z)\) is generally poorly constrained by data with sometimes inconsistent measurements \citep[\eg][]{10.1111/j.1365-2966.2008.13114.x}. We discuss in \S~\ref{sec:discussion} where this model breaks down, \eg a double power law during reionization or temperature fluctuations arising from an inhomogeneous reionization \citep{2019MNRAS.486.4075O, 2019MNRAS.490.3177W}, and the resulting implications for dark matter bounds.

\subsubsection{\label{sec:IGM_input}Ionization state}

The ionization state of each species is determined by the balance between the UVB ionization rates and recombination effects. Uncertainty in the ionization rates is degenerate in the flux power spectrum with the mean amount of absorption in the transmitted flux or the effective optical depth \(\tau_\mathrm{eff}\). We therefore marginalize over the uncertainty in the ionization state by varying \(\tau_\mathrm{eff}\); we parameterize \(\tau_\mathrm{eff}\) at each redshift by a multiplicative correction \(\tau_0 (z)\) to a fiducial model \citep{2019ApJ...872..101B}:
\begin{equation}\label{eq:tau_0}
\tau_\mathrm{eff} (z) = \tau_0 (z) \times 1.4 \times 10^{-3} \times (1 + z)^4.
\end{equation}

\subsubsection{\label{sec:IGM_input}Simulation input and output}

The thermal and ionization history in our hydrodynamical simulations (see \S~\ref{sec:sims}) is determined by a set of input spatially-uniform UV ionization and heating rates for each primordial species [HI, HeI, HeII] as a function of redshift; our default set of rates comes from Ref.~\cite{2012ApJ...746..125H}\footnote{This assumes the IGM to be optically thin; we discuss in \S~\ref{sec:discussion} where this model breaks down, \eg the impact of radiative transfer \cite[\eg][]{2015MNRAS.453.2943C}, as well as the implications for dark matter bounds.}. In order to generate a wide range of IGM histories, we vary the default photo-heating rates \(\epsilon_{0,i}\) by an amplitude \(H_\mathrm{A}\) and an overdensity \(\Delta\)--dependent rescaling \(H_\mathrm{S}\) such that heating rates \(\epsilon_i = H_\mathrm{A} \epsilon_{0,i} \Delta^{H_\mathrm{S}}\) are used. This uniformly scales the heating rates with redshift. In order to vary the redshift dependence, we employ the reionization model of Ref.~\cite{2017ApJ...837..106O}. We vary two free parameters of the model: the redshift of hydrogen reionization \(z_\mathrm{rei}\), defined as the redshift at which the volume-averaged ionization fraction in HII \(= 0.5\); and, for the first time in studying nCDM models using the flux power spectrum, the total heat injection during hydrogen reionization \(T_\mathrm{rei}\). The reionization model of Ref.~\cite{2017ApJ...837..106O} self-consistently generates heating and ionization rates for chosen reionization parameters, only transitioning to the default rates once ionization is complete. The model, in particular, removes spurious heating at high redshift before and during reionization, which has been present in previous work \citep[\eg][]{1996ApJ...461...20H, 2001cghr.confE..64H, Faucher_Gigu_re_2009, 2012ApJ...746..125H}.

We vary \(H_\mathrm{A} \in [0.05, 3.5]\), \(H_\mathrm{S} \in [-1.3, 0.7]\), \(z_\mathrm{rei} \in [6, 15]\), \(T_\mathrm{rei} \in [1.5 \times 10^4, 4 \times 10^4]\,\mathrm{K}\). In order to be further agnostic about the IGM states over which we marginalize and to have as robust nCDM bounds as possible, in our emulator and likelihood function (see \S~\ref{sec:method_nCDM}), we allow IGM parameters to vary independently with redshift. To achieve this, we label each simulation at each redshift with the parameter set \([\tau_0, T_0, \widetilde{\gamma}, u_0]\). This no longer restricts the redshift evolution of our IGM model to that determined by the input model alone, which may be incomplete \citep[see, \eg other estimations of UVB heating and ionization rates;][]{2019MNRAS.485...47P}. In particular, this allows us to account for uncertainty in HeII reionization, which may start heating the IGM as early as \(z \sim 4.5\), without explicitly varying helium parameters in the input model. This would otherwise further extend the dimensionality of the model, which has important implications regarding the construction of the emulator (see \S~\ref{sec:method_nCDM}).

At each redshift we restrict our sample values of \([T_0, \widetilde{\gamma}, u_0]\) to those generated in our simulations; \ie we do not employ the practice of rescaling the temperature-density relation after a simulation snapshot has been saved \citep[\eg][]{2019ApJ...872..101B, 2019arXiv190507410T}. This practice may be inaccurate when using the smallest scales in the flux power spectrum, \eg as rescaling the temperatures of simulation particles ignores their peculiar velocities. The effect of this on the flux power spectrum may not be fully captured by other IGM parameters (\ie \(u_0\)). Future work will assess the impact of this. However, we do employ the common practice of rescaling mock spectrum optical depths in order to increase our sampling of \(\tau_\mathrm{eff}\); Ref.~\cite{2015MNRAS.446.3697L} tested the accuracy of this procedure. This process is relatively computationally cheap and means that, for each simulation snapshot, we can have an arbitrary number of effective optical depth values; we discuss this further in \S~\ref{sec:method_nCDM}.

\subsubsection{\label{sec:cosmo}Cosmology}

The flux power spectrum at these redshifts (\(z \sim 5\)) and scales (\(k_\mathrm{f} \sim 10^{-2} - 10^{-1}\,\mathrm{s}\,\mathrm{km}^{-1}\)) is sensitive to aspects of the cosmological model, which must be marginalized for robust nCDM bounds. In particular, to search for a cut-off in the power spectrum arising from an nCDM model, we marginalize over the amplitude \(A_\mathrm{s} \in [1.2 \times 10^{-9}, 2.5 \times 10^{-9}]\) and slope \(n_\mathrm{s} \in [0.9, 0.995]\) of the primordial power spectrum. We note that these differ from the standard cosmological parameters \citep{2018arXiv180706205P} as we define them at a pivot scale \(k_\mathrm{p} = 2\,\mathrm{Mpc}^{-1}\), which lies in the range of scales we are considering. This requires translating CMB priors from the larger pivot scale used by the Planck Collaboration. In the data analysis we present in Ref.~\cite{2020RogersPRL}, we use \textit{Planck}-derived priors on the primordial power spectrum. It follows that during the acquisition of training points during Bayesian optimization (see \S~\ref{sec:method_ULA}), these are largely drawn from that prior distribution (through the exploitation term) and so are clustered around a \textit{Planck} fiducial cosmology (see Figs.~\ref{fig:emulator} and \ref{fig:posterior}). We also vary in the initial set of simulations, the fractional matter energy density \(\Omega_\mathrm{m} \in [0.26, 0.33]\), although in our initial optimization analysis presented below, we fix it to a fiducial value from \textit{Planck} \citep{2018arXiv180706205P, refId0}, \(\Omega_\mathrm{m} = 0.3209\). All other cosmological parameters are fixed to fiducial values from the \textit{Planck} 2018 results \citep{refId0}, in particular, \(\Omega_\mathrm{b} h^2 = 0.022126\) and \(\Omega_\mathrm{c} h^2 = 0.12068\).

\subsection{\label{sec:method_nCDM}Gaussian process emulation and likelihood function}

In order to infer bounds on nCDM parameters and to marginalize over the nuisance parameters identified in \S~\ref{sec:model_cosmo}, we want to be able to model data as a function of theory parameters. A challenge arises because each evaluation of our theory vector naively requires the calculation of a hydrodynamical simulation (see \S~\ref{sec:sims}) and numerical methods of inference (\eg Markov-chain Monte Carlo sampling; MCMC) may require millions of these evaluations. This would be computationally infeasible given the numerical requirements for sufficiently converged simulations (see Appendix \ref{sec:appendix_convergence}). We follow the solution we presented in Refs.~\cite{2019JCAP...02..031R, 2019JCAP...02..050B}, which is to construct a ``surrogate model'' or ``emulator'' for the simulations using a Gaussian process \citep[\eg][]{rasmussen2003gaussian} and to ensure convergence and sufficient accuracy in its construction by the method of Bayesian emulator optimization. A Gaussian process emulator is sufficiently computationally-cheap to evaluate that standard inference methods like MCMC can be used.

We emulate separately at each redshift \(z = [4.2, 4.6, 5.0]\), the flux power spectrum as a function of the nCDM parameters (\S~\ref{sec:model_nCDM}) and the nuisance IGM and cosmological parameters (\S~\ref{sec:model_cosmo}): \(\vec{\theta} = [\alpha, \beta, \gamma, \tau_0 (z = z_i), T_0 (z = z_i), \widetilde{\gamma} (z = z_i), u_0 (z = z_i), n_\mathrm{s}, A_\mathrm{s}, \Omega_\mathrm{m}]\). As discussed in \S~\ref{sec:sims}, we emulate the flux power spectrum at 45 sample wavenumbers in inverse-distance units (\(h\,\mathrm{Mpc}^{-1}\)). This accounts for the fact that the velocity wavenumbers of a given simulation depend on \(\Omega_\mathrm{m}\) (by the Hubble law) and so emulating in velocity space would be complicated by having simulation modes binned differently at each training point. In the likelihood function, we then linearly interpolate the flux power spectrum to velocity wavenumber bins.

\subsubsection{\label{sec:GP}Gaussian process}

Our emulator model is a Gaussian process; this is a stochastic process such that a finite number of simulation outputs (\ie the flux power spectrum) forms a multivariate Gaussian distribution: \(P_\mathrm{f} (\vec{\theta}) \sim \mathcal{N}(0, K(\vec{\theta}, \vec{\theta'}; \vec{\psi}))\). Here, for clarity, we have dropped in the notation the dependence of \(P_\mathrm{f}\) on \(k_\mathrm{f}\) and \(z\). The free hyper-parameters \(\vec{\psi}\) of the emulator model for \(P_\mathrm{f} (\vec{\theta})\) will be optimized (or ``trained'') using a set of ``training simulations'' \(P_\mathrm{f} (\vec{\theta'})\) evaluated at a set of points \(\vec{\theta'}\). The zero mean condition is approximated by normalizing the flux power spectra by the median value in the training set. Having chosen the training set of simulations (see \S~\ref{sec:training_sims}), the final part of the model to determine is the covariance kernel \(K(\vec{\theta}, \vec{\theta'}; \vec{\psi})\). We use a linear combination of a squared exponential kernel \big[or radial basis function; RBF; \(\sigma_\mathrm{RBF}^2 \exp\left(-\frac{(\vec{\theta} - \vec{\theta'})^2}{2 l^2}\right)\)\big], a linear kernel [\(\sigma_\mathrm{linear}^2 \vec{\theta}.\vec{\theta'}\)] and a constant noise kernel [\(\sigma_\mathrm{noise}^2\)]; \ie \(\vec{\psi} = [\sigma_\mathrm{RBF}^2, \sigma_\mathrm{linear}^2, \sigma_\mathrm{noise}^2, l]\). The Gaussian process model constitutes a wide prior in function space and so allows emulation of the flux power spectrum without any strong prior knowledge of its parameter dependence. Moreover, the model is probabilisitic and so any prediction of the flux power spectrum away from the training points will have an associated uncertainty which can be propagated to our statistical model. The choice of covariance kernel is motivated by wanting a flexible model (hence the squared exponential), recognizing that linear interpolation has been a reasonable approach in previous efforts \citep[\eg][]{2017PhRvD..96b3522I} (hence the linear kernel) and having a term which can account for noise in the training data (although we find that this is always negligible).

\subsubsection{\label{sec:training_sims}Training simulations}

In the construction of the training data, we adapt the methods we presented in Refs.~\cite{2019JCAP...02..031R, 2019JCAP...02..050B} to this more general parameterization. We run an initial set of fifty simulations spanning the wide prior range on the nCDM, IGM and cosmological parameters set out above (\S~\ref{sec:model_nCDM} and \ref{sec:model_cosmo}). To achieve this, we choose the parameter values of these simulations from a Latin hypercube sampling scheme \citep{10.2307/1268522}. This sampling scheme is chosen because it has good space-filling properties with a low number of sampling points. It distributes \(N\) training points such that along each parameter axis, the full space is sampled in \(N\) equally-divided sub-spaces. We generate many different hypercubes and use the one that maximizes the minimum Euclidean distance between samples. This Latin hypercube is constructed for our input simulation parameters, \ie \([\alpha, \beta, \gamma, H_\mathrm{A}, H_\mathrm{S}, z_\mathrm{rei}, T_\mathrm{rei}, n_\mathrm{s}, A_\mathrm{s}, \Omega_\mathrm{m}]\), since we do not know \textit{a priori} the mapping from input to output IGM parameters. This means that in our initial set of training simulations, the IGM parameters \([T_0 (z = z_i), \widetilde{\gamma} (z = z_i), u_0 (z = z_i)]\) are not sampled by a Latin hypercube. We find, however, that these parameters are still evenly sampled (see Fig.~\ref{fig:emulator} for the distribution of training points) and the subsequent application of Bayesian emulator optimization will correct for any loss of precision. A partial degeneracy appears between \(T_0 (z = z_i)\) and \(u_0 (z = z_i)\) such that snapshots with high \(T_0\) and low \(u_0\) (and \textit{vice versa}) are not produced. This constitutes a physical prior on these parameters as very hot IGMs with little previous heating (and \textit{vice versa}) would be contrived. We therefore set a prior probability distribution that excludes parameter space outside the convex hull of the training points in this plane (at each redshift). The full details of the prior distribution specific to the data analysis we present in Ref.~\cite{2020RogersPRL} can be found in that reference.

We also note that \(\tau_0 (z = z_i)\) does not form part of the Latin hypercube since (as discussed in \S~\ref{sec:model_cosmo}), we can have an arbitrary number of \(\tau_0\) samples by rescaling mock spectrum optical depths. For each training simulation, we have ten values of \(\tau_0 (z = z_i) \in [0.75, 1.25]\) (\ie allowing for 25\% corrections to the fiducial model). These are evenly distributed such that across all the training points, we have \(10\,N = 500\) different values of \(\tau_0\) for maximal support in this dimension.

The framework we propose is as follows: the initial base nCDM emulator can be used for parameter inference on the many particular nCDM models it can characterize (see \S~\ref{sec:model_nCDM}). This means that the process of Bayesian emulator optimization (the iterative addition of training simulations until parameter inference has converged) should be applied for each particular set of dark matter model parameters (and nuisance IGM/cosmological parameters). In \S~\ref{sec:emu_ULA}, we demonstrate the example of optimizing the emulator for inferring a bound on the mass of the ultra-light axion. Other dark matter model parameters can be inferred by equivalent analyses. We therefore defer a discussion of the specifics of the Bayesian emulator optimization to \S~\ref{sec:method_ULA}.

\subsubsection{\label{sec:emu_training}Emulator training}

We now discuss how a given emulator (whether optimized or not) propagates to the final parameter inference. With a given training set, we optimize (or ``train'') the hyperparameters \(\vec{\psi}\) of the Gaussian process by maximizing the marginal likelihood of the training data. We train each redshift separately but use the same set of optimized hyperparameters for each wavenumber within each redshift bin. We do not model correlations between different wavenumbers; Ref.~\cite{2019arXiv190507410T} found this to be a sufficient approximation. We can then make predictions for the flux power spectrum at arbitrary values of \(\vec{\theta^*}\) by evaluating the posterior predictive distribution (PPD) for the flux power spectrum conditional on the training data. This is a Gaussian distribution with mean \(\vec{\mu} = K_* K^{-1} P_\mathrm{f}(\vec{\theta'})\) and variance \(\vec{\sigma}^2 = K_{**} - K_* K^{-1} K_*^\mathrm{T}\), where \(K_* = K(\vec{\theta^*}, \vec{\theta'}; \vec{\psi})\) and \(K_{**} = K(\vec{\theta^*}, \vec{\theta^*}; \vec{\psi})\). The generic behaviour of the Gaussian process PPD is that its uncertainty increases as a function of distance to the nearest training points. It follows that more precise theoretical estimates of the flux power spectrum will in general be made in regions of parameter space which are more densely sampled by training data; the Bayesian emulator optimization exploits this feature of Gaussian processes (see \S~\ref{sec:method_ULA}).

\subsubsection{\label{sec:likelihood_func}Likelihood function}

In this analysis, we use a Gaussian likelihood function, although this initial emulator could be used with any likelihood function that can robustly propagate the theoretical uncertainty arising from the Gaussian process emulation. The theory vector is taken to be the PPD mean \(\vec{\mu}(\vec{\theta})\) and the uncertainty in the flux power spectrum emulation is propagated by adding in quadrature the diagonal PPD covariance \(\vec{\sigma}^2(\vec{\theta})\) to the data covariance.

\section{\label{sec:emu_ULA}Ultra-light axion emulator}

In this section, we present the construction of the Bayesian-optimized emulator for inferring a bound on the mass of ultra-light axions as the dark matter using the Lyman-alpha forest flux power spectrum. This emulator is built using the base nCDM emulator presented in \S~\ref{sec:emu_nCDM}. It constitutes an example of how to construct an optimized emulator for a particular dark matter model characterized by the nCDM parameters. Other dark matter candidates could be studied in an equivalent way. In \S~\ref{sec:model_ULA}, we explain the ultra-light axion model we employ; and in \S~\ref{sec:method_ULA}, we discuss the method of Bayesian emulator optimization and how this connects to estimation of the posterior probability distribution.

\subsection{\label{sec:model_ULA}Ultra-light axion model}

\begin{figure}
\includegraphics[width=\columnwidth]{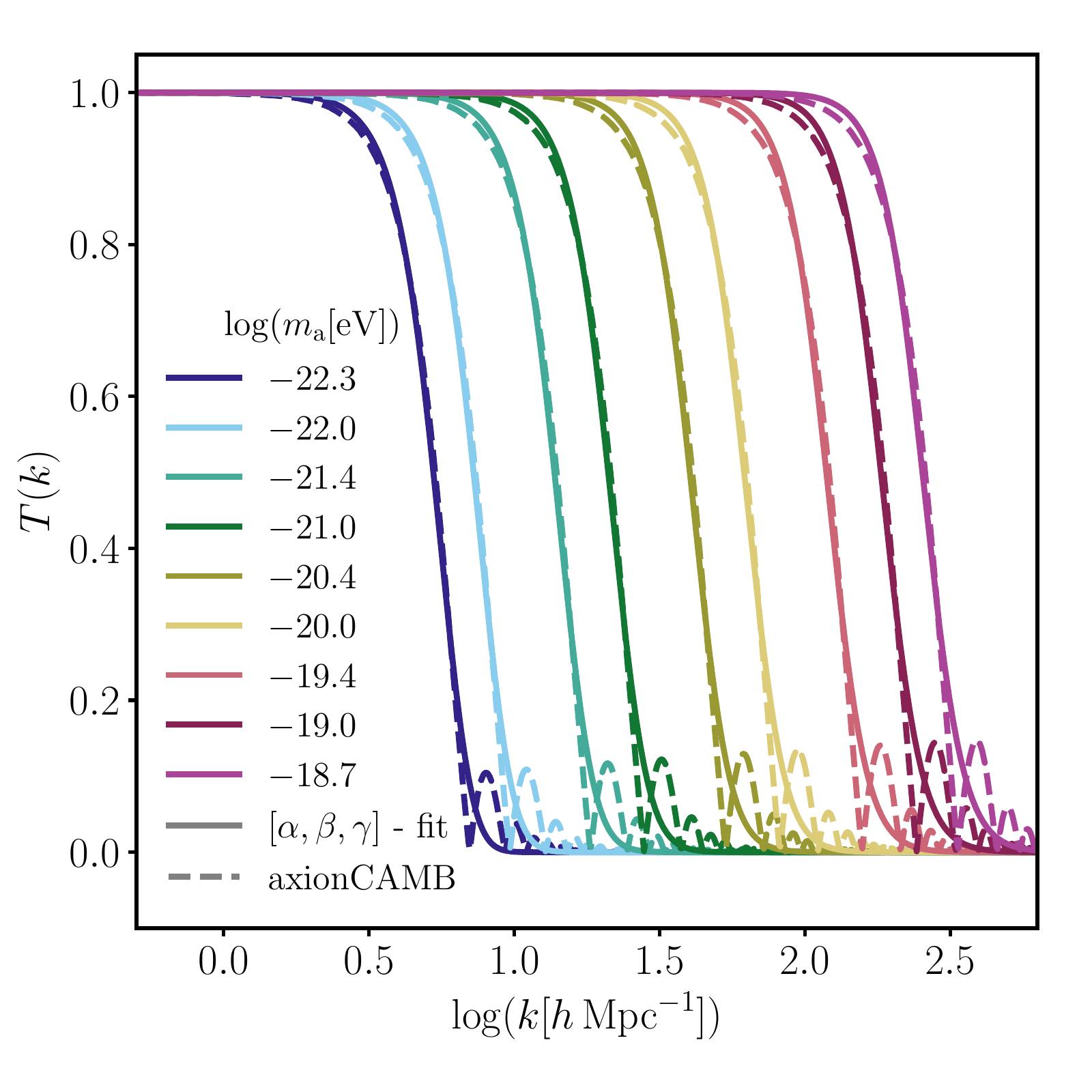}
\caption{\label{fig:transfer_ULA}The transfer function \(T(k)\) (as defined in Eq.~\eqref{eq:transfer} at \(z = 99\)) for the ultra-light axion dark matter model as a function of comoving distance wavenumber \(k\) and axion mass \(m_\mathrm{a}\). The dashed lines show \(T(k)\) as calculated by the Boltzmann code \texttt{axionCAMB} \citep{2017PhRvD..95l3511H}; the solid lines show our nCDM model fit using the \([\alpha, \beta, \gamma]\) parameters. The nCDM model cannot characterize the small-scale oscillations in the \texttt{axionCAMB} transfer functions but in the power spectrum, these are strongly suppressed relative to the initial cut-off scale.}
\end{figure}

\begin{figure}
\includegraphics[width=\columnwidth]{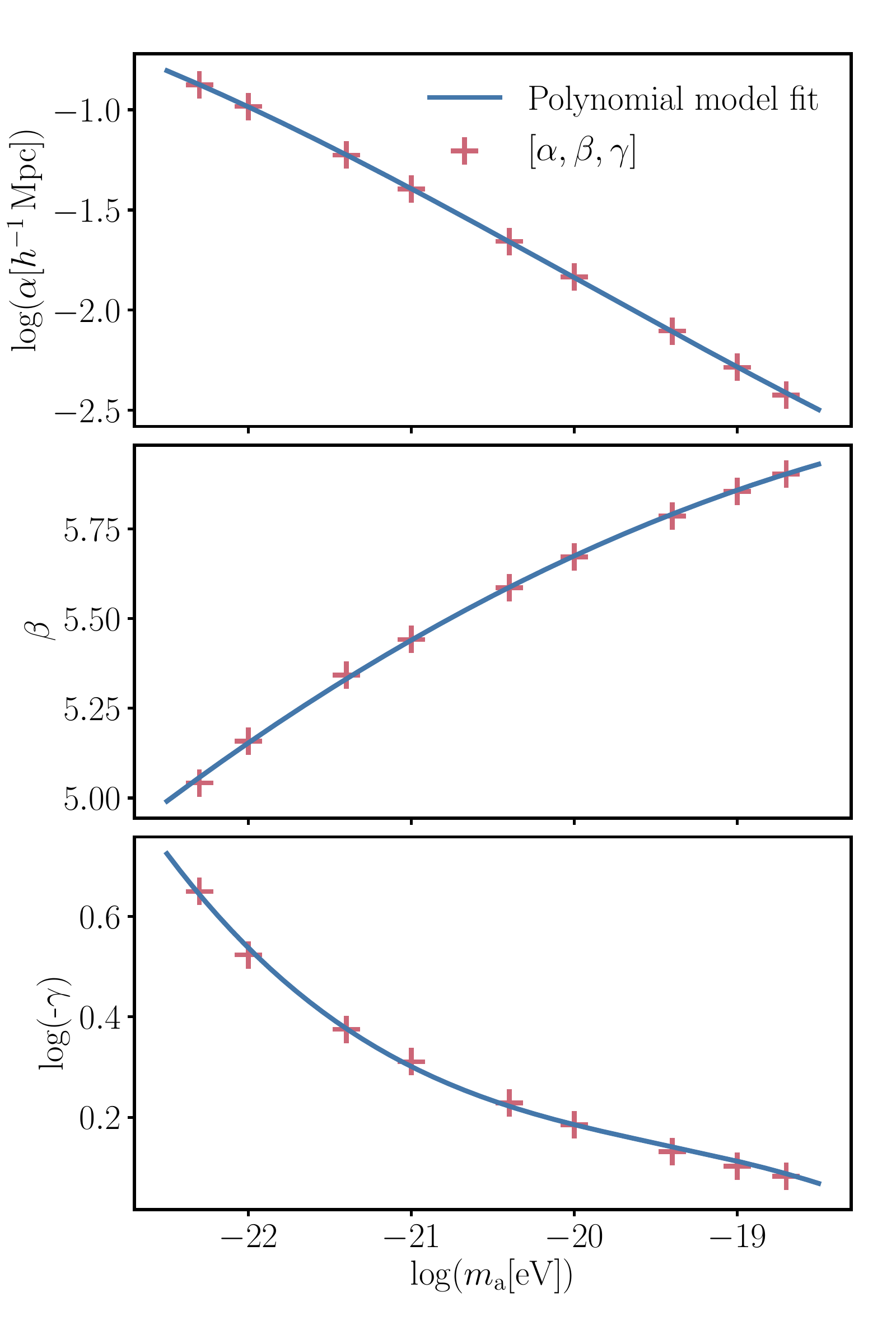}
\caption{\label{fig:transfer_ULA_poly}The nCDM model fit for ultra-light axion dark matter transfer functions as a function of axion mass \(m_\mathrm{a}\). \textit{From top to bottom}, the three nCDM parameters \([\alpha, \beta, \gamma]\), showing a polynomial model fit (see text for more details).}
\end{figure}
Figures \ref{fig:transfer_ULA} and \ref{fig:transfer_ULA_poly} demonstrate the construction of our ultra-light axion dark matter model. We fit \citep[using the Levenberg-Marquardt algorithm;][]{levenberg1944method, doi:10.1137/0111030} the nCDM parameters \([\alpha, \beta, \gamma]\) to the ULA transfer functions \citep[as calculated by the Boltzmann code \texttt{axionCAMB};][]{2017PhRvD..95l3511H} as a function of axion mass \(m_\mathrm{a}\). We use polynomial functions to model the dependences of \([\alpha, \beta, \gamma]\) on \(m_\mathrm{a}\).\footnote{\(\log(\alpha [h^{-1}\,\mathrm{Mpc}]) = 5.5 \times 10^{-3} \mathcal{M}^3 + 0.33 \mathcal{M}^2 + 6.2 \mathcal{M} + 33.1\); \(\beta = -0.026 \mathcal{M}^2 - 0.82 \mathcal{M} - 0.45\); \(\log(- \gamma) = -0.013 \mathcal{M}^3 - 0.75 \mathcal{M}^2 - 14.7 \mathcal{M} - 9.6\); where \(\mathcal{M} = \log(m_\mathrm{a} [\mathrm{eV}])\).} We fit the \(\alpha\) dependence at the fiducial value of \(h = 0.6686\) \citep{refId0}, to which \(h\) is fixed in this part of the analysis (see \S~\ref{sec:method_ULA}). No particular physical meaning should be ascribed to the polynomial model we have employed; it serves as a simple mapping from axion mass to our nCDM parameters. The polynomial model strictly only applies within our prior bounds on axion mass, in which our model parameters are fit, for \(\log(m_\mathrm{a} [\mathrm{eV}]) \in [-22, -19]\).

Our ground truth for the ULA transfer functions is given by \texttt{axionCAMB}\footnote{\url{https://github.com/dgrin1/axionCAMB}.}, a modified version of the Einstein-Boltzmann solver \texttt{CAMB} \citep{Lewis:2002ah}. It solves the coupled Friedmann/Klein-Gordon system of equations for a homogeneous ULA field in an expanding universe. At early times, when \(m_\mathrm{a} \ll a \mathcal{H}\) (\(a\) is the cosmological scale factor; \(\mathcal{H}\) is the conformal Hubble parameter), the evolving axion equation of state \(w(a)\) and adiabatic sound speed are used to evolve axion energy and pressure perturbations along with the other cosmological energy components. At late times, when \(m_\mathrm{a} \gg a \mathcal{H}\), the equations become stiff and the axion field is well-approximated as a pressureless fluid (\(w = 0\)) and the perturbations are evolved in the WKB approximation for ULAs with scale-dependent sound speed \citep[\eg][]{2010PhRvD..82j3528M}. This approximation captures the suppression of small-scale axion perturbations, observed as the sharp cut-off in the transfer functions in Fig.~\ref{fig:transfer_ULA}, without requiring the resolution of the short ULA oscillation timescale (\(\sim m_\mathrm{a}^{-1}\)). We find that the nCDM model (Eq.~\eqref{eq:transfer}) can well characterize the power spectrum cut-off produced by \texttt{axionCAMB} (Fig.~\ref{fig:transfer_ULA}). It is not able to characterize the suppressed small-scale oscillations in the transfer function. However, even in the initial conditions of our simulations (\S~\ref{sec:sims}) in the matter power spectrum, these oscillations will be further suppressed since \(P(k) \propto T^2(k)\). They will be erased further again in propagation to our observable, the flux power spectrum at \(z \sim 5\) by non-linearities in the density field and the projection from a 3D to a 1D line-of-sight power spectrum. Considering that current data measure the flux power spectrum at the precision of \(\sim 10 \% - 25 \%\) \citep[\eg][]{2019ApJ...872..101B}, we conclude that our bounds on axion mass will be insensitive to this approximation and driven by the non-detection of the initial sharp cut-off in the power spectrum. We will discuss this further and compare to the approximations in previous efforts in \S~\ref{sec:discussion}. We consider only the case where ULAs form the entirety of the dark matter; future analyses can extend our parameter space to consider sub-dominant contributions to the dark matter energy density (see \S~\ref{sec:discussion}).

\subsection{\label{sec:method_ULA}Bayesian-optimized emulator and posterior distribution}

In order to infer a bound on the mass of ultra-light axions as the entirety of the dark matter (and marginalize over nuisance IGM/cosmological parameters; see \S~\ref{sec:model_cosmo}), we must numerically sample the posterior probability distribution for the full set of inference parameters \(\vec{\phi} = [\log(m_\mathrm{a} [\mathrm{eV}]), \tau_0 (z = z_i), T_0 (z = z_i), \widetilde{\gamma} (z = z_i), u_0 (z = z_i), n_\mathrm{s}, A_\mathrm{s}]\), for \(z_i = [4.2, 4.6, 5.0]\). We use the Gaussian likelihood function presented in \S~\ref{sec:method_nCDM}.

Although \(\Omega_\mathrm{m}\) is varied in our emulator (\S~\ref{sec:model_cosmo}), since relevant data do not constrain this parameter, here we fix \(\Omega_\mathrm{m} = 0.3209\), which is equivalent to \(h = 0.6686\) \citep{refId0}. We estimate the posterior probability distribution (with a given prior distribution) using the affine-invariant MCMC ensemble sampler \texttt{emcee} \citep{2013PASP..125..306F}.

In order to evaluate the theory flux power spectrum, we use the nCDM emulator described in \S~\ref{sec:emu_nCDM}. We map from \(\log(m_\mathrm{a} [\mathrm{eV}])\) to \([\alpha, \beta, \gamma]\) using the model constructed in \S~\ref{sec:model_ULA}. However, as mentioned in \S~\ref{sec:method_nCDM}, the initial base emulator with fifty training simulations is un-optimized and not guaranteed to give converged estimates of the posterior distribution. This can arise when the uncertainty in emulator prediction (the variance in its posterior predictive distribution) dominates over the data error (covariance) at the peak of the true posterior distribution (which is not \textit{a priori} known). This can lead to weakened parameter constraints (as the emulator error broadens the peak of the likelihood) or even bias determination of the maximum posterior point due to the non-stationarity of the emulator variance. We account for this problem by optimizing the emulator training set (iteratively adding to it) such that the true posterior peak is determined; this is the Bayesian emulator optimization we introduced in Refs.~\cite{2019JCAP...02..031R, 2019JCAP...02..050B}.

At each iteration of the Bayesian emulator optimization, after training the emulator on the existing training set and evaluating the posterior distribution by MCMC (see above), an acquisition function is maximized in order to determine the position in parameter space for the next training simulation. Following Ref.~\cite{2019JCAP...02..031R}, we use the modified GP-UCB (Gaussian process upper confidence bound) acquisition function \citep{Cox97sdo:a, auer2002using, auer2002finite, dani2008stochastic} \(\mathcal{A} (\vec{\widetilde{\phi}}) = \mathcal{P}(\vec{\widetilde{\phi}} | \vec{d}) + \widetilde{\alpha} \vec{\sigma}^\mathrm{T} (\vec{\widetilde{\phi}}) \Sigma^{-1} \vec{\sigma} (\vec{\widetilde{\phi}})\). \(\mathcal{P}(\vec{\widetilde{\phi}} | \vec{d})\) is the natural logarithm of the posterior probability given data \(\vec{d}\); \(\Sigma\) is the data covariance matrix; and \(\vec{\sigma} (\vec{\widetilde{\phi}})\) is the emulator error vector (or standard deviation of the Gaussian process PPD; see \S~\ref{sec:method_nCDM}). Following Refs.~\cite{2009arXiv0912.3995S, 2010arXiv1012.2599B}, we set the acquisition hyperparameter \(\widetilde{\alpha} = 0.85\); this is roughly equivalent to having \(\sim 1 \sigma\) confidence in estimation of the posterior distribution during optimization. Since, as discussed in \S~\ref{sec:emu_nCDM}, we can have an arbitrary number of training points in the \(\tau_0 (z = z_i)\) dimensions due to a computationally-cheap simulation post-processing, we do not optimize the training set in these dimensions. In the acquisition function, \(\vec{\widetilde{\phi}}\) is the truncated set of inference parameters without \(\tau_0 (z = z_i)\). It can be seen that the acquisition function has two terms which balance (in the first term) \textit{exploitation} of approximate knowledge of our objective function (the posterior distribution) and (in the second term) \textit{exploration} of the full prior volume. The exploitation term prefers adding training simulations at the peak of the posterior (which is the most important region to characterize accurately). The exploration term prefers adding simulations where uncertainty in the estimation of the true (zero emulator error) posterior is large, which is equivalent to regions where the ratio of emulator to data covariance is large. The linear combination balances these two components. In order to prevent the exploitation term getting stuck in local or spurious maxima of the true posterior, the exploration term and the stochastic element in the acquisition (see below) are vital. It is possible to vary the balance between exploitation and exploration using the acquisition hyperparameter.

We find the maximum of the acquisition function using a truncated Newton algorithm \citep{doi:10.1137/0721052}. In order to ensure we explore the full peak of the posterior distribution (\ie the \(95 \%\) to \(99 \%\) credible region), we then choose the position of the next training simulation by adding a random displacement (drawn from a multivariate Gaussian distribution) to the maximum acquisition point \citep{2017arXiv170400520J}. The size of this displacement is tuned to the size of the credible region which we want to characterize accurately. A further subtlety arises since the acquisition function is defined on the space of the (truncated) inference parameters, which include the output IGM parameters \([T_0, \widetilde{\gamma}, u_0]\). In order to determine which input IGM parameters to use for the next training simulation, we map back to \([H_\mathrm{A}, H_\mathrm{S}, z_\mathrm{rei}, T_\mathrm{rei}]\) using radial basis function interpolation between the existing training set \citep{buhmann_dyn_1993}. This method is related to, but less general than, the Gaussian process emulation and serves as a simple yet robust method for this mapping. Once the next training simulation is run, the \(\tau_0 (z = z_i)\) samples are redistributed such that the \(10 N\) (\(N\) is the total number of training simulations including the optimization set) points evenly sample the full prior range. The emulator is then re-trained (the emulator hyperparameters \(\vec{\psi}\) are re-optimized) and the posterior distribution is re-estimated by MCMC using the new iteration of the emulator.

The process of Bayesian emulator optimization continues and training simulations are added to the emulator until convergence is observed. We use improved ``double-lock'' convergence criteria. As in Ref.~\cite{2019JCAP...02..031R}, one criterion is that successive estimations of the posterior distribution (\eg as summarized by marginalized statistics like the mean and \(1 \sigma\) and \(2 \sigma\) constraints) do not change within a given tolerance; we show the results of this convergence test in \S~\ref{sec:results} and Fig.~\ref{fig:convergence}. The second criterion is that the exploration term of the acquisition function evaluated at the parameter positions of optimization simulations converges towards zero. This ensures that the acquisition of new training data is then dominated by exploitation, \ie that training points are being added at the peak of the posterior and that the ratio of emulator to data covariance is tending towards zero in this region. The results of this convergence test are shown in \S~\ref{sec:results} and Fig.~\ref{fig:exploration}.

In this analysis, we use the batch version of Bayesian emulator optimization \citep{2019JCAP...02..031R}. This involves, at each iteration, proposing a batch of optimization simulations at once, running these in parallel and then re-training the emulator only after the entire batch has been added to the training set. The first simulation in a batch of a given size is chosen as detailed above. Subsequent members of a batch are chosen in a similar way, except that the output (flux power spectra) from previous members are not yet known. However, the positions in parameter space for these members are known and this information can at least update the emulator error distribution, since the PPD variance depends only on the positions of training data and not the data themselves (see \S~\ref{sec:method_nCDM}). This partial information updates the exploration term of the acquisition function for iterations within a batch. Batch optimization can be preferred to balance efficient parallel use of computational resources with the loss of optimality in the acquisition of new training data. In this study, we start with three batches of five and then switch to batches of two, although we note that the third and sixth batches are short of one simulation each, owing to hardware issues that led to an incomplete calculation.

\section{\label{sec:results}Tests}

In this section, we conduct tests of convergence (\S~\ref{sec:convergence}) and cross-validation (\S~\ref{sec:validation}) on the Bayesian-optimized ULA emulator presented in \S~\ref{sec:emu_ULA}.

\subsection{\label{sec:convergence}Tests of convergence}

\begin{figure*}
\includegraphics[width=1.3\columnwidth]{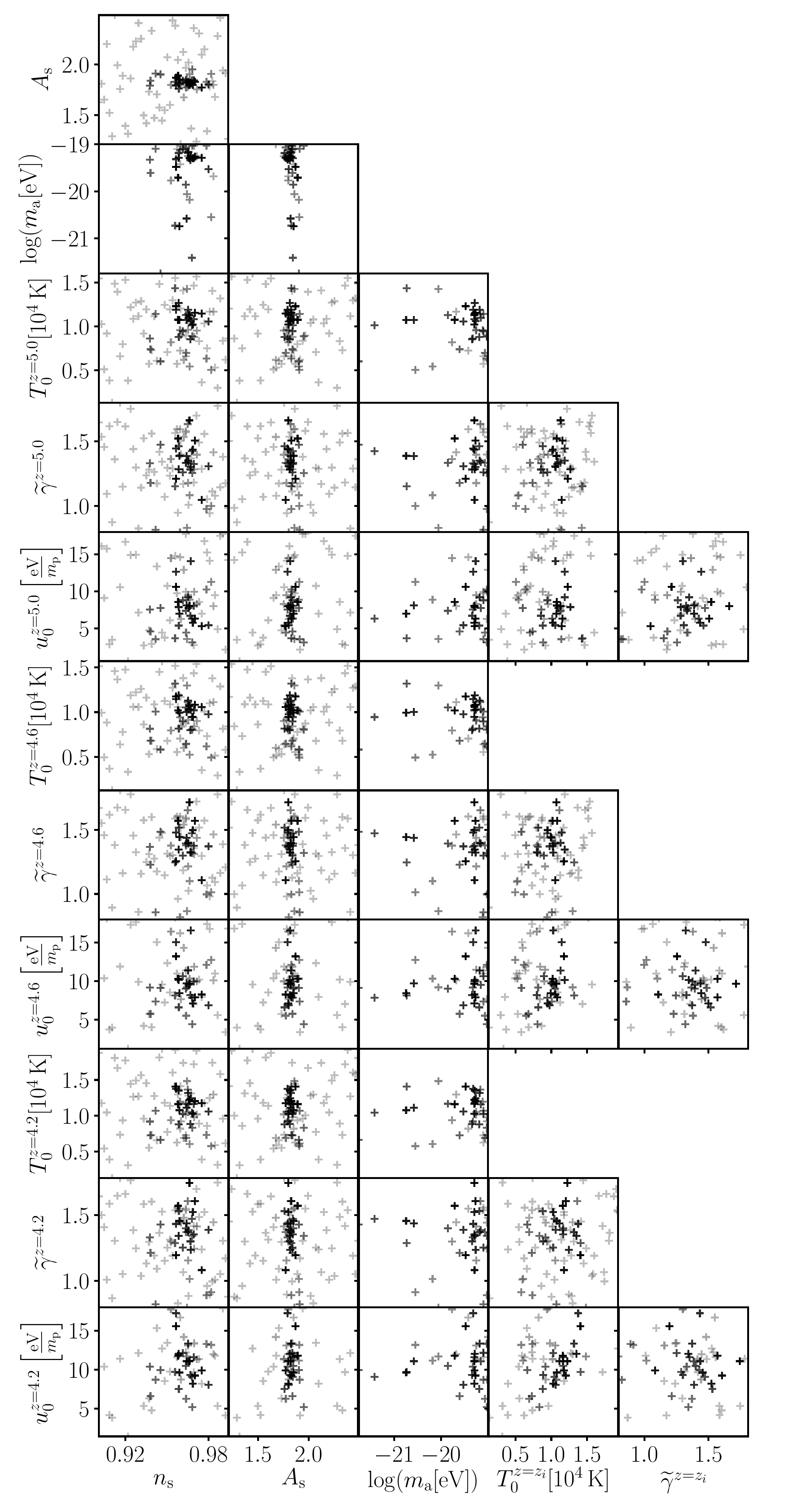}
\caption{\label{fig:emulator}The positions of the Bayesian-optimized ULA emulator training simulations projected onto each parameter plane. The shade of the crosses indicates the iteration at which the training point is added, such that the initial set is shaded lightest and the final optimization point is shaded darkest. We show only the planes where parameters are emulated together (\ie not the planes with parameters from different redshift bins; or the \(\tau_0 (z = z_i)\) planes which are densely sampled by simulation post-processing, see \S~\ref{sec:sims}). It follows that on the \(x\) axis, \(z_i\) corresponds to the redshift indicated on the \(y\) axis. We note that for \(\log (m_\mathrm{a} [\mathrm{eV}])\), the initial set fully spans the \([\alpha, \beta, \gamma]\) volume and does not project onto that axis; however, the initial set contributes to the final emulator model as we always emulate in \([\alpha, \beta, \gamma]\). It can be seen how the darker-shaded optimization points concentrate in a smaller sub-volume, which corresponds to the peak of the posterior distribution (see Fig.~\ref{fig:posterior}).}
\end{figure*}
\begin{figure*}
\includegraphics[width=\textwidth]{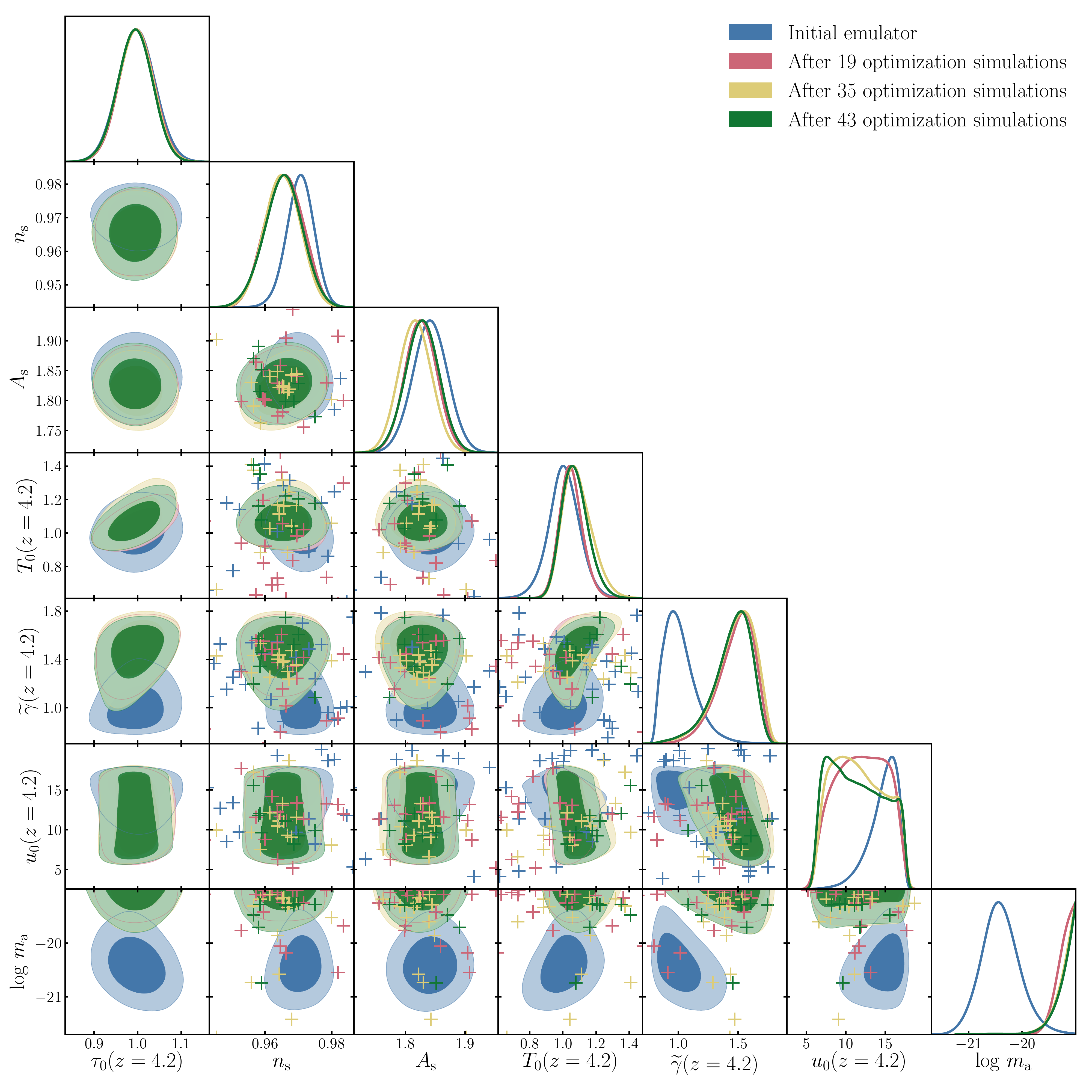}
\caption{\label{fig:posterior}The convergence of the posterior distribution (set of marginalized 1D and 2D distributions) using the Bayesian-optimized ULA emulator. Each set of colored contours shows the estimate of the posterior at various iterations of the emulator. The darker and lighter shaded regions respectively indicate the \(68 \%\) and \(95 \%\) credible regions. In each plane (except the \(\tau_0 (z = 4.2)\) planes which are densely sampled by simulation post-processing, see \S~\ref{sec:sims}), crosses indicate the projected positions of emulator training simulations. The crosses are color-coded by the iteration of the emulator by which the training points are added. We note that for \(\log (m_\mathrm{a} [\mathrm{eV}])\), the initial set fully spans the \([\alpha, \beta, \gamma]\) volume and does not project onto that axis; however, the initial set contributes to the final emulator model as we always emulate in \([\alpha, \beta, \gamma]\).} \(T_0 (z = 4.2)\) is in units of \(10^4\,\mathrm{K}\); \(u_0 (z = 4.2)\) is in units of \(\mathrm{eV}\,m_\mathrm{p}^{-1}\), where \(m_\mathrm{p}\) is the proton mass; and \(m_\mathrm{a}\) is in units of eV. For clarity, we show only the projections onto the IGM parameters at \(z = 4.2\); we observe similar convergence at the other redshifts we consider.
\end{figure*}

Figure \ref{fig:emulator} indicates the distribution of emulator training points, projected onto each parameter plane. It can be seen that the initial Latin hypercube set of fifty (see \S~\ref{sec:emu_nCDM}; most lightly shaded in Fig.~\ref{fig:emulator}) fully spans the parameter space. The optimization simulations (43 in total; see \S~\ref{sec:emu_ULA}; more darkly shaded in Fig.~\ref{fig:emulator}) concentrate in a smaller sub-space, which corresponds to the peak of the posterior distribution (see Fig.~\ref{fig:posterior} and below). There is however a degeneracy between \(T_0 (z = z_i)\) and \(u_0 (z = z_i)\) since it is unphysical to generate simulations with high \(T_0\) and low \(u_0\) and \textit{vice versa}. We reiterate that these are not samples from the posterior distribution and that the posterior is always estimated with fully converged MCMC chains.

\begin{figure}
\includegraphics[width=\columnwidth]{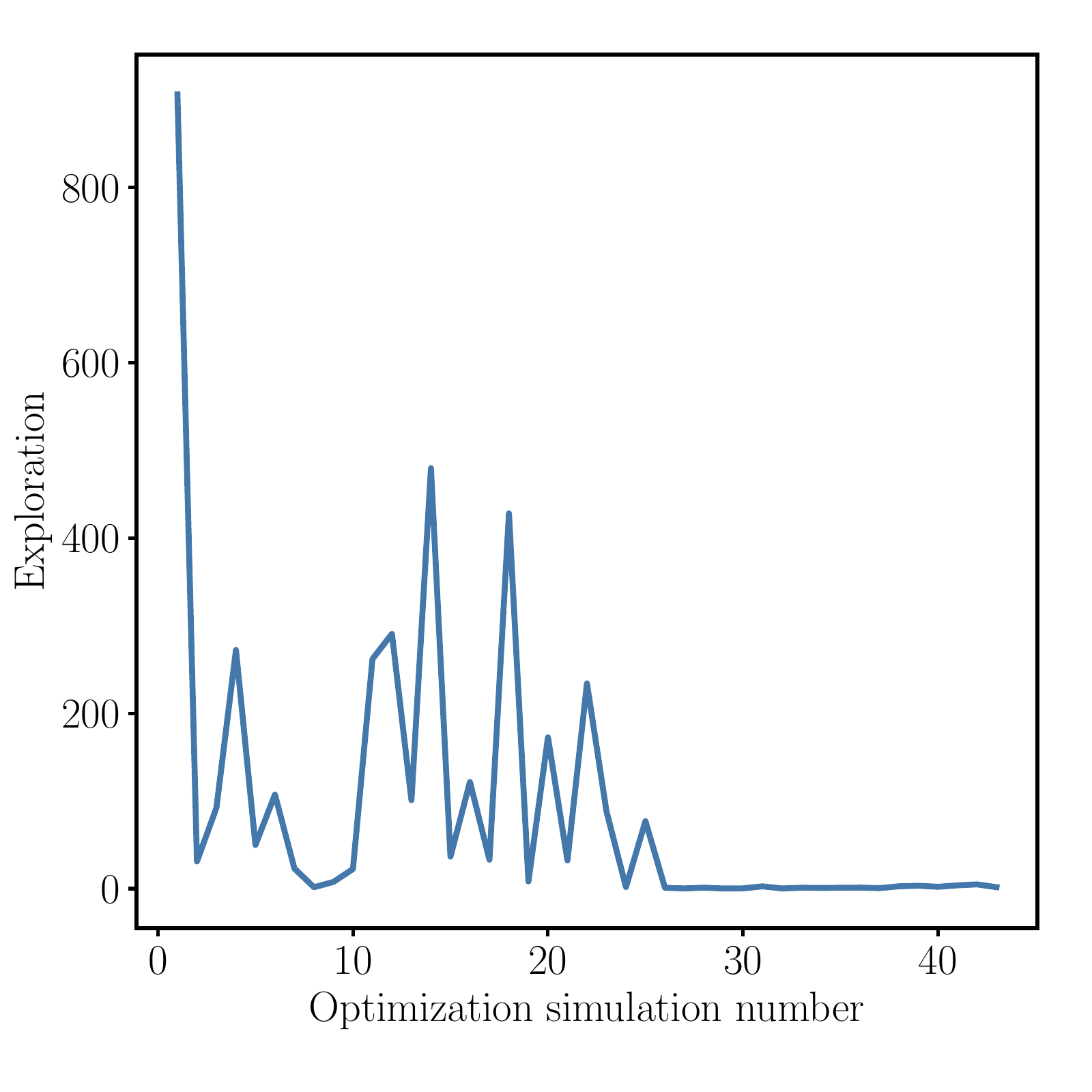}
\caption{\label{fig:exploration}The convergence of the exploration term of the acquisition function (see \S~\ref{sec:method_ULA}) as a function of optimization simulation number for the Bayesian-optimized ULA emulator.}
\end{figure}
Figure \ref{fig:exploration} shows the first of the two convergence tests we use. It indicates how the exploration term of the Bayesian optimization acquisition function (see \S~\ref{sec:method_ULA}) tends towards zero as more training simulations are added to the emulator. The exploration term is effectively the ratio of emulator to data covariance. Its convergence to zero reflects how the optimization becomes more dominated by exploitation, \ie the proposal of training simulations is at the peak of the posterior and the ratio of emulator to data covariance is negligible there. This means that estimation of the posterior peak (by MCMC) tends towards the true posterior, with negligible propagated emulator error.

\begin{figure*}
\includegraphics[width=\textwidth]{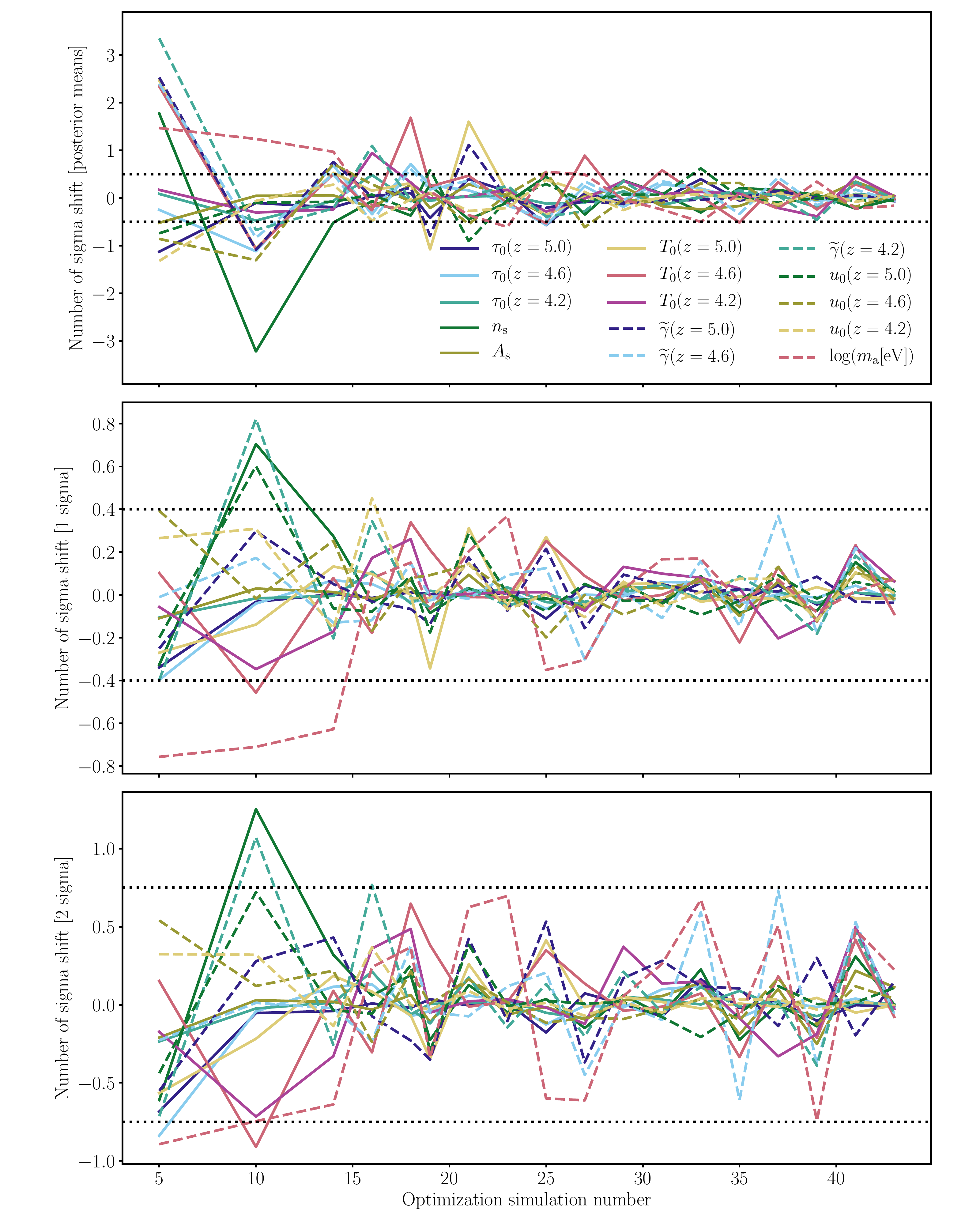}
\caption{\label{fig:convergence}The convergence of summary statistics of the posterior distribution using the Bayesian-optimized ULA emulator. \textit{From top to bottom}, the number of sigma shift (defined by the marginalized posteriors at a given optimization epoch) between emulator iterations for the 1D marginalized posterior means, \(1 \sigma\) and \(2 \sigma\) constraints. Each colored line shows the convergence for each model parameter.}
\end{figure*}
Figure \ref{fig:posterior} indicates the convergence of the posterior distribution (see \S~\ref{sec:method_ULA}) as summarized by the set of marginalized 1D and 2D distributions. For clarity, we show only the projections in the IGM parameters at \(z = 4.2\); we observe similar convergence at the other redshifts that we consider. It can be seen in each projection that the posterior estimation converges as more optimization simulations are added to the emulator training set. In most projections, there is a significant shift from using the initial emulator (\S~\ref{sec:emu_nCDM}) to having added 19 optimization points. In particular, the axion mass \(m_\mathrm{a}\) goes from being (spuriously) detected to setting a lower bound. However, after this point, the shifts in the posterior are much less significant; and following the iteration after which 35 optimization points have been added, the shifts in the posterior are statistically negligible. We track this more quantitatively using summaries of the marginalized posterior (marginalized mean and \(1 \sigma\) and \(2 \sigma\) constraints).

Figure \ref{fig:convergence} shows for each model parameter, the number of sigma shift (defined by the marginalized posteriors at a given optimization epoch) in these quantities between iterations as a function of optimization simulation number. It can be seen that by the end of the optimization, the shifts are small, indicating the desired robustness in the estimation of the posterior distribution. In Fig.~\ref{fig:posterior}, in the planes of parameters that are emulated together (remembering that different redshifts are emulated separately), we indicate by colored crosses the projected positions of optimization simulations. It can be seen that by Bayesian optimization, the optimization simulations are more concentrated in the posterior peak than the initial set which is evenly distributed in the full prior volume. This reduces the emulator error within the posterior peak, leading to accurate determination of the flux power spectrum (see \S~\ref{sec:validation}) and hence the likelihood function in this region.

\subsection{\label{sec:validation}Tests of cross-validation}

\begin{figure*}
\includegraphics[width=\textwidth]{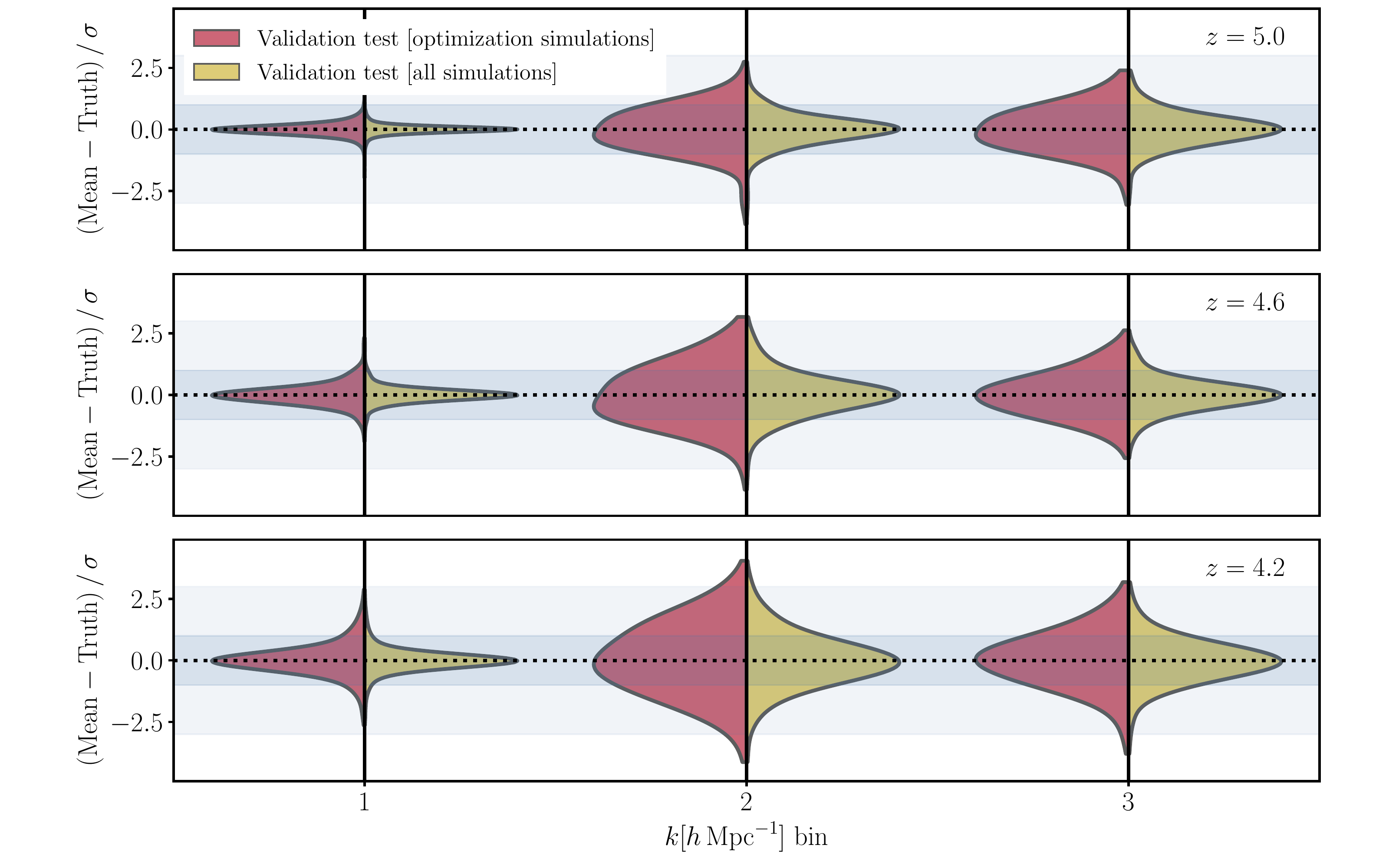}
\caption{\label{fig:validation}A violin plot (set of kernel density estimation histograms) showing the ratio of empirical (emulator mean -- truth) to predicted emulator error for leave-one-out cross-validation simulations, (\textit{from left to right}) as a function of binned wavenumber \(k\) and (\textit{from top to bottom}) redshift \(z\). The left set of PDFs considers only the simulations added by Bayesian optimization; the right set considers all simulations including the initial set. For comparison, we indicate \(1 \sigma\) and \(3 \sigma\) limits respectively by the darker and lighter shaded bands. If the emulator model was fitting perfectly, each distribution would be a unit Gaussian; the distribution being fatter indicates under-fitting (underestimating emulator error) and the distribution being narrower indicates over-fitting (overestimating emulator error). The wavenumber bins span the following ranges (where \(k\) is in \(h\,\mathrm{Mpc}^{-1}\)): 1 [\(0.6 < k < 10\)]; 2 [\(10 < k < 19\)]; 3 [\(19 < k < 28.3\)].}
\end{figure*}
We carry out leave-one-out cross-validation of the Bayesian-optimized emulator. This involves, in turn, removing one of the training simulations, re-training the reduced emulator and then comparing its prediction to the truth of the validation simulation that was removed\footnote{This is actually leave-\textit{ten}-out cross-validation since we remove in turn all ten effective optical depth samples (at each redshift) associated with a given training simulation. We do not remove optical depth samples one-by-one since this would be a trivial test considering that these dimensions can be densely sampled by a simulation post-processing step. The number of optical depth samples was tested in Ref.~\cite{2019JCAP...02..050B}.}. We remove simultaneously the training simulation from each redshift bin since these are emulated separately. Figure \ref{fig:validation} shows the distribution of the ratio of empirical-to-predicted emulator error in the flux power spectrum at the validation simulations, as a function of wavenumber \(k\) and redshift \(z\). The empirical error is the difference between the emulator PPD mean (see \S~\ref{sec:method_nCDM}) and the true flux power spectrum as measured from the validation simulation. The predicted emulator error is the standard deviation of the emulator PPD. If the emulator model is fitting well, the distribution of the error ratio should be a unit Gaussian. When the distribution is fatter than this, this indicates under-fitting, \ie that the emulator error is being underestimated and that the Gaussian process model does not have sufficient flexibility to describe accurately the training or the validation data. When the distribution is narrower than a unit Gaussian, this indicates over-fitting, \ie that the emulator error is being overestimated and that the Gaussian process has too much flexibility and is fit too closely to the training data. This has negative implications for the generalizability of the emulator model for unseen (\eg validation) data.

In Fig.~\ref{fig:validation}, the distributions are split into two sets: the left-handed halves of each violin consider the validation test only for the simulations added by Bayesian optimization, while the right-handed halves consider all the simulations including the initial Latin hypercube. In each case, we consider only the final, converged emulator and in cross-validation, optimize the emulator on all other 92 training simulations. In splitting the distributions that are shown, we indicate the two populations that are formed. When the initial set of simulations is included, the overall distribution tends towards overestimating the emulator error, in particular on the largest scales (smallest \(k\)). However, when considering only the optimization simulations (and hence the training points at the peak of the posterior distribution), the Gaussian process fit is better, although the error is still overestimated on the largest scales; there is some mild underestimation on intermediate scales. Overestimating the emulator error will tend to weaken parameter constraints as large errors broaden the peak of the posterior.

\begin{figure*}
\includegraphics[width=\textwidth]{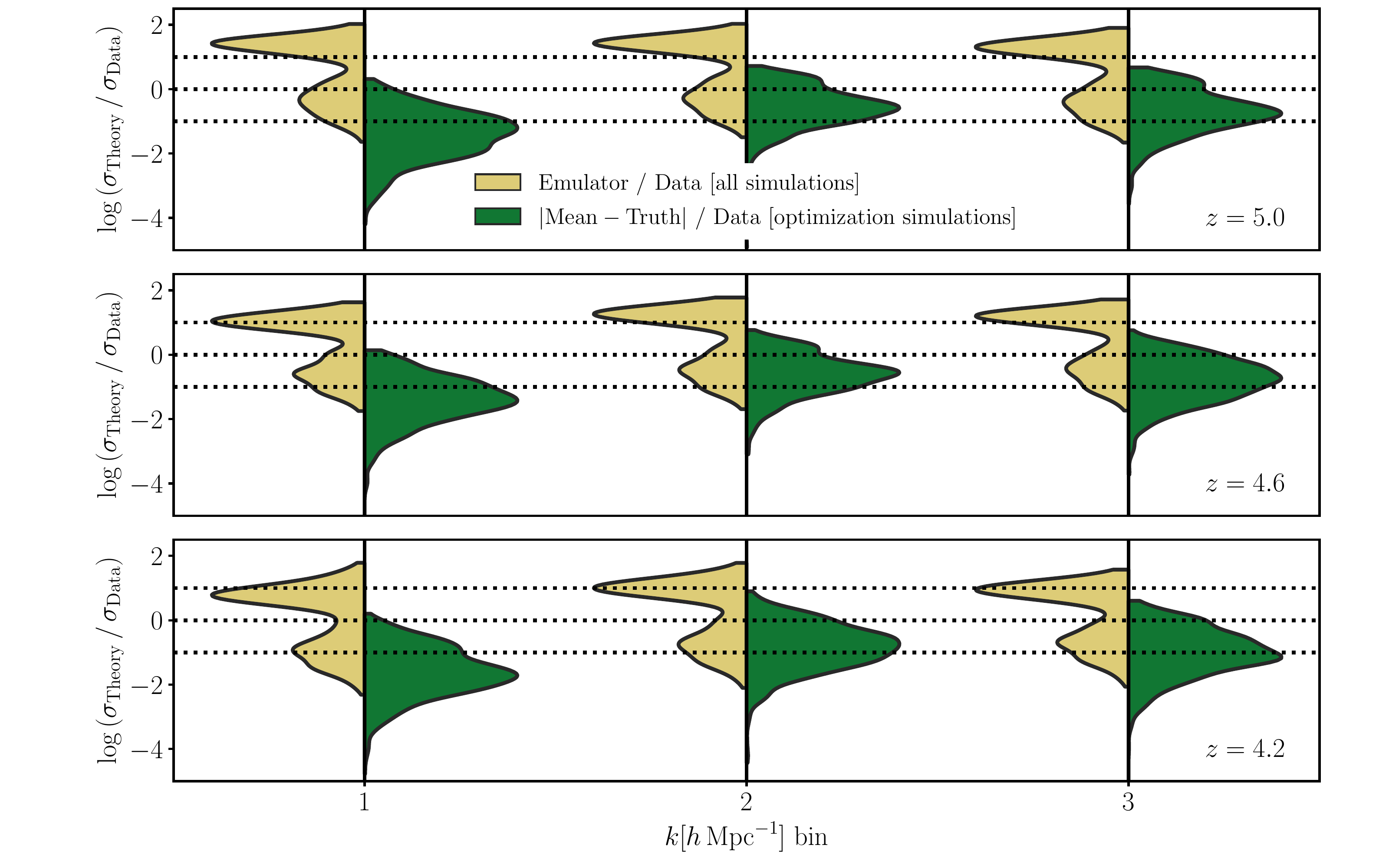}
\caption{\label{fig:data_error}A violin plot (set of kernel density estimation histograms) showing (the logarithm of) the ratio of theory to data error in the flux power spectrum, (\textit{from left to right}) as a function of binned wavenumber \(k\) and (\textit{from top to bottom}) redshift \(z\). \textit{For each element}, the left PDF shows the ratio of emulator error to data error; for the majority of training data in the optimization set (which constitutes the lower-valued mode in each case), this is \(< 1\). Since Fig.~\ref{fig:validation} shows through cross-validation a tendency to overestimate the prediction uncertainty, the right PDFs show the ratio of the empirical error (for validation simulations only in the optimization set for clarity) to data error. This ratio is often \(\ll 1\) (and less than the theoretical emulator-to-data error ratio, see also Fig.~\ref{fig:validation}), indicating no statistically-significant bias in emulation of the flux power spectrum. See more details in the text. The wavenumber bins span the following ranges (where \(k\) is in \(h\,\mathrm{Mpc}^{-1}\)): 1 [\(0.6 < k < 10\)]; 2 [\(10 < k < 19\)]; 3 [\(19 < k < 28.3\)].}
\end{figure*}
Figure \ref{fig:data_error} compares the empirical and predicted emulator error (\ie respectively the numerator and denominator in the ratio illustrated in Fig.~\ref{fig:validation}) to the data error (this is the square root of the diagonal elements of the data covariance used in our data analysis presented in Ref.~\cite{2020RogersPRL}). The distributions of (the logarithm of) the ratio of emulator-to-data error are in general bi-modal; the lower-valued modes come from the optimization simulations, while the other modes come from the initial Latin hypercube. This reflects how the ratio of emulator-to-data error is significantly lower for the optimization simulations (and therefore at the peak of the posterior distribution) than for the initial Latin hypercube (in fact, for clarity, we only show the lower-valued optimization set mode for the empirical ratios in the right-hand PDFs). This is a direct consequence (and desired outcome) of the Bayesian optimization, which builds up training information at the posterior peak and locally reduces emulator uncertainty in the flux power spectrum, which leads to more accurate determination of the peak of the posterior distribution. The left-handed and right-handed distributions in each violin show the ratio of emulator-to-data error respectively for the predicted emulator error (PPD standard deviation) and the empirical emulator error (difference between PPD mean and truth). As is consistent with the test in Fig.~\ref{fig:validation}, the empirical ratio is consistently shifted to lower values than the predicted ratio, reflecting how the emulator tends towards over-fitting the training data and overestimating the emulator uncertainty.

Nonetheless, for the optimization simulations, the mode average of the predicted emulator-to-data error ratio is always less than one. This means that in the posterior peak (\(\sim 95 \%\) credible region), uncertainty in the flux power spectrum (\ie the diagonal of the covariance matrix in the likelihood function; see \S~\ref{sec:method_nCDM}) is dominated by uncertainty in the data. As discussed in Ref.~\cite{2019JCAP...02..031R}, this is a key requirement for converged estimation of the posterior distribution as it means that any residual uncertainty in the emulator model is not statistically significant. Further, a key motivation for the Bayesian emulator optimization that we use is to move beyond relying on heuristic assumptions on the accuracy requirements of an emulator. The tests of convergence that we present in \S~\ref{sec:convergence} provide a stringent test that the addition of more training points (and so further reducing the emulator error) makes no statistically significant change to the inferred parameter constraints. Considering the tendency to overestimate the emulator error discussed above (in particular on the largest scales), the ratio of empirical emulator-to-data error is often \(< 0.1\). This shows that there is no statistically significant bias (relative to the data uncertainty and accounting for the theoretical emulator error included in the likelihood function) in the emulation of the flux power spectrum.

\begin{figure*}
\includegraphics[width=\textwidth]{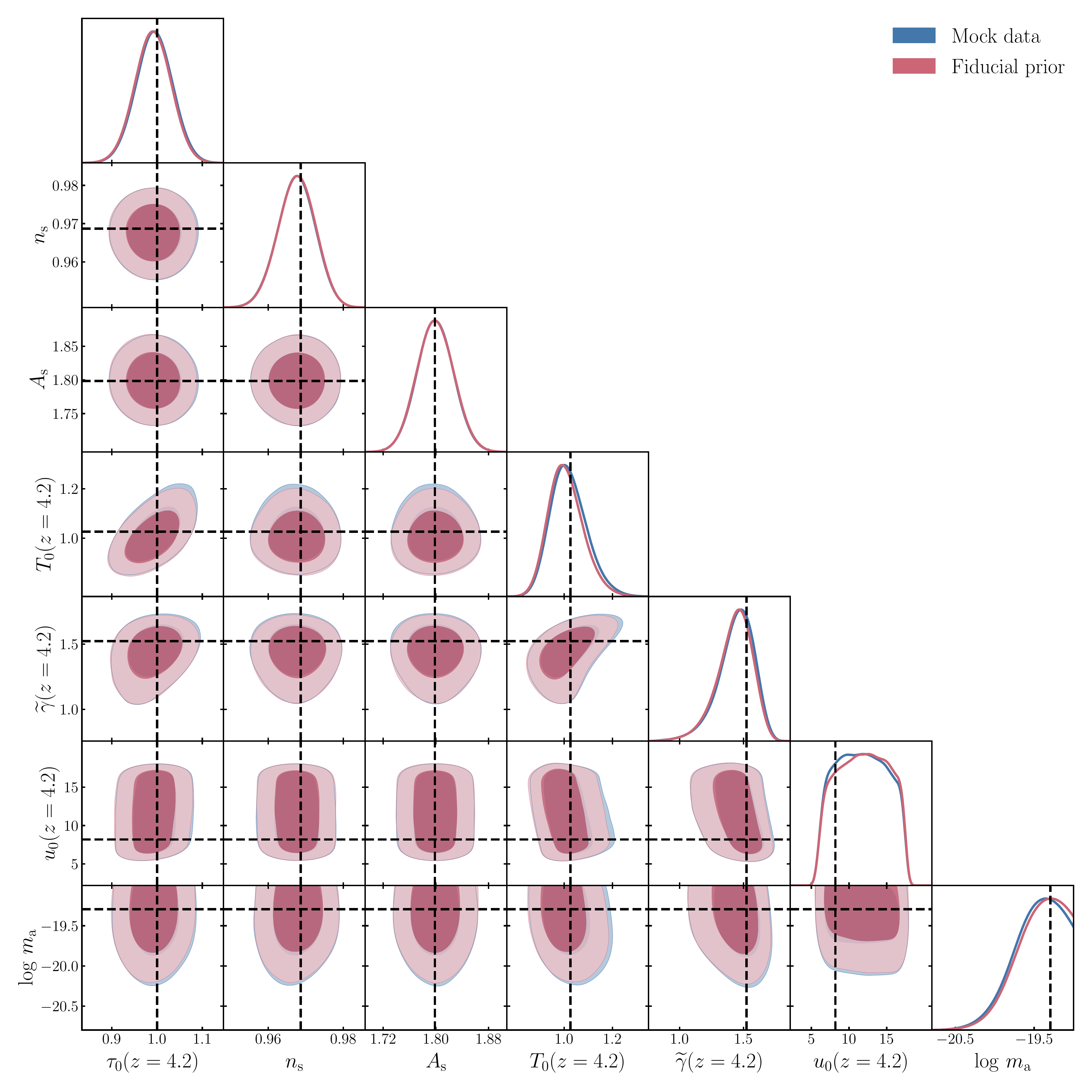}
\caption{\label{fig:posterior_mock}Cross-validation test on the inferred posterior distribution using the Bayesian-optimized ULA emulator: set of marginalized 1D and 2D distributions given mock simulated data removed from the training set. The darker and lighter shaded regions respectively indicate the \(68 \%\) and \(95 \%\) credible regions. The blue set of contours (obscured by the red set) uses a Gaussian prior on \(T_0 (z = z_i)\) with means matched to the true values of the mock data (and standard deviations of 3000 K): respectively at \(z = [5.0, 4.6, 4.2]\), [10208, 9471, 10262] K. As a test of sensitivity to the prior distribution, the red set of contours instead uses \(T_0 (z = z_i)\) prior means set to the fiducial simulation values from our data analysis in Ref.~\cite{2020RogersPRL}: respectively at \(z = [5.0, 4.6, 4.2]\), [8022, 7651, 8673] K. Dashed lines indicate the true parameter values of the mock data. \(T_0 (z = 4.2)\) is in units of \(10^4\,\mathrm{K}\); \(u_0 (z = 4.2)\) is in units of \(\mathrm{eV}\,m_\mathrm{p}^{-1}\), where \(m_\mathrm{p}\) is the proton mass; and \(m_\mathrm{a}\) is in units of eV. For clarity, we show only the projections onto the IGM parameters at \(z = 4.2\); we observe similarly unbiased inference at the other redshifts we consider.}
\end{figure*}

Figure \ref{fig:posterior_mock} shows a test of cross-validation on the inferred posterior distribution using the Bayesian-optimised ULA emulator. We show the (marginalized) posterior distribution on the ULA model and nuisance IGM and cosmological parameters given mock simulated data, which have been removed from the emulator training set and the emulator re-trained on the reduced set. This is a test as to whether parameter values can be inferred without bias using an emulator for the flux power spectrum that has not seen the mock data. We see that the truth is almost perfectly recovered at the maximum (marginalized) posterior values (we expect no scatter from the mock data as we do add any noise realization) and always within a fraction of a sigma. The same is observed for the nuisance IGM parameters at other redshifts, which we do not show for clarity in the figure. We do not re-optimize (by Bayesian optimization) the emulator for the mock data, and the mock data are taken from the optimization set of the training simulations (those added by Bayesian optimization as shown in \S~\ref{sec:convergence}).

As a test of the sensitivity of the parameter inference to the prior distribution, we vary the prior on \(T_0 (z = z_i)\). The blue set of contours (obscured by the red set) uses a Gaussian prior on \(T_0 (z = z_i)\) with means matched to the true values of the mock data (and standard deviations of 3000 K; see caption). The red set of contours instead uses \(T_0 (z = z_i)\) prior means set to the fiducial simulation values from our data analysis in Ref.~\cite{2020RogersPRL} (also see caption). We see almost no resulting shift in the posterior.

\section{\label{sec:discussion}Discussion}

\subsection{\label{sec:discuss_DM}Dark matter model}

The model we employ to describe the effect of non-cold dark matter on the matter power spectrum \citep[first introduced in][]{2017JCAP...11..046M} can characterize many different dark matter candidates with different physical mechanisms suppressing the small-scale power spectrum. These include, \eg warm dark matter with a free streaming scale \citep[\eg][]{2005PhRvD..71f3534V} or dark matter interacting with itself or baryons \citep[\eg][]{2000PhRvL..84.3760S, 2014PhRvD..89b3519D}; this is discussed further in Ref.~\cite{2017JCAP...11..046M}. For the nCDM model to apply, the detectable effect on the power spectrum relative to the cold dark matter limit must be fully captured by modified initial conditions at \(z \sim 100\). We consider here and in an accompanying data analysis \citep{2020RogersPRL}, the example of ultra-light axion dark matter (\S~\ref{sec:emu_ULA}). We fit the nCDM model to ULA matter power spectra calculated by the modified Boltzmann code \texttt{axionCAMB} \citep[][see \S~\ref{sec:model_ULA}]{2017PhRvD..95l3511H}. We find the fit to be good aside from suppressed small-scale power spectrum oscillations which cannot be captured by the nCDM model. We argue, however, that given the precision of flux power spectrum measurements on the relevant scales \citep[\(\sim 10 \%\) to \(25 \%\);][]{2019ApJ...872..101B}, these suppressed oscillations are currently undetectable and so the nCDM model captures the relevant phenomenology. This compares to the approximate ULA power spectrum transfer function employed in previous Lyman-alpha forest studies \citep{2017PhRvL.119c1302I}, which is a semi-analytical approximation presented in Ref.~\cite{2000PhRvL..85.1158H}. Although this model (imperfectly) captures small-scale oscillations (which do not persist to the flux power spectrum at \(z \sim 5\)), the initial cut-off in the power spectrum is sharper than for the full calculation made by \texttt{axionCAMB}. We find that the nCDM model better captures the relevant cut-off in the power spectrum (see Fig.~\ref{fig:transfer_ULA}).

\subsection{\label{sec:discuss_IGM}Intergalactic medium model}

Robust dark matter bounds using the Lyman-alpha forest must marginalize over uncertainty in the thermal and ionization state of the intergalactic medium, which leads to a similar, but distinguishable, suppression in the small-scale flux power spectrum (see \S~\ref{sec:model_cosmo}). We model the IGM as obeying a single power-law temperature-density relation (TDR; Eq.~\eqref{eq:TDR}). We also argue that pressure smoothing can be tracked by the cumulative energy deposited per unit mass at the mean density \(u_0 (z)\); and exploit the fact that the strength of the photo-ionizing UV background is degenerate in the flux power spectrum with the effective optical depth of Lyman-alpha absorption. This compares to previous Lyman-alpha forest studies of dark matter \citep[\eg][]{2017PhRvL.119c1302I, 2017PhRvD..96b3522I} which have usually tracked the pressure smoothing by varying the redshift of hydrogen reionization. When varying this parameter simultaneously with the TDR at each redshift, this can lead to marginalizing over unphysical IGMs where the energy deposited by (re)ionization is inconsistent with the instantaneous gas temperature. This may spuriously weaken dark matter bounds. In order to avoid this, we model the IGM using only ``output'' parameters that describe the state of the IGM at a fixed time-slice. In the data analysis presented in Ref.~\cite{2020RogersPRL}, we prevent unphysical redshift evolution of the IGM by preventing changes in \(T_0\) and \(u_0\) between adjacent redshift bins greater than a conservative threshold value. We also improve the input for our simulations by employing the reionization model of Ref.~\cite{2017ApJ...837..106O}. Compared to previous prescriptions \citep[\eg][]{1996ApJ...461...20H, 2001cghr.confE..64H, Faucher_Gigu_re_2009, 2012ApJ...746..125H}, it removes spurious high-redshift heating and provides additional physical flexibility by varying both the timing of and total heat injection during reionization.

However, an underlying assumption in our (and previous) IGM models is that reionization is a spatially homogeneous process. This is an approximation, as ionization and heating are expected to expand in overlapping bubbles around the (uncertain) sources of ionizing radiation \citep[\eg][]{1999ApJ...514..648M}. The resultant temperature and ionizing background spatial fluctuations that can persist to the IGM at \(z \sim 5\) \citep[\eg][]{2016MNRAS.460.1328D, 2019MNRAS.486.4075O} have been suggested as a possible source of systematic error in Lyman-alpha forest dark matter bounds \citep[\eg][]{2017PhRvD..95d3541H}. Ref.~\cite{2019MNRAS.490.3177W} found, using simulations which couple gas hydrodynamics with radiative transfer \citep{2019MNRAS.485..117K}, that the relevant effect on the flux power spectrum is below the sensitivity of current data \citep[\eg][]{2019ApJ...872..101B}. Nonetheless, as modeling and data improve, future versions of the emulator can benefit from explicitly marginalizing over the effect of inhomogeneity in both hydrogen and helium reionizations \citep[\eg][]{2016MNRAS.460.1328D, 2019MNRAS.486.4075O, Suarez:2017xqg}.

\section{\label{sec:concs}Conclusions}

We have presented a general framework for testing the nature of dark matter using cosmological data and \(N\)-body simulations, specifically implemented for searching for deviation from cold dark matter using the Lyman-alpha forest. It consists of a Gaussian process emulator, trained using hydrodynamical simulations to predict how the Lyman-alpha forest flux power spectrum responds to different dark matter models, the state of the intergalactic medium and the spectrum of primordial fluctuations. This is necessary since MCMC methods for inference would otherwise be impractical owing to the computational cost of \(N\)-body simulations. The emulator exploits a flexible parameterization of the effect of non-cold dark matter on the matter power spectrum \citep[introduced by][]{2017JCAP...11..046M}. This can accurately capture the suppression in the power spectrum arising from many different dark matter models (relative to the cold dark matter limit), \eg warm \citep[\eg][]{2005PhRvD..71f3534V} or interacting \citep[\eg][]{2000PhRvL..84.3760S, 2014PhRvD..89b3519D} dark matter or ultra-light axion dark matter \citep{2000PhRvL..85.1158H}, among others \citep[see][]{2017JCAP...11..046M}. The emulator can be optimized for the analysis of particular dark matter models by the Bayesian emulator optimization we presented in Ref.~\cite{2019JCAP...02..031R}. This tests for convergence in the fidelity of the emulator model by adaptively building up the emulator's training information.

We demonstrate the example of constructing a Bayesian-optimized emulator for the mass of ultra-light axion dark matter. We carry out tests of convergence with respect to the number of hydrodynamical simulations input to the emulator; and cross-validation for the accuracy of flux power spectrum prediction. These tests confirm the robustness of the data analysis we carry out in Ref.~\cite{2020RogersPRL}. The emulator framework we present here will facilitate the future analysis of other dark matter candidates by an equivalent procedure. We argue that the application of Bayesian optimization to emulators of cosmological ``effective theories,'' where many different models are described by a single set of equations, can be a powerful approach for robust and computationally-efficient parameter inference from the cosmic large-scale structure.

\begin{acknowledgments}
KKR thanks Jose O{\~n}orbe for sharing his code for the reionization model of Ref.~\cite{2017ApJ...837..106O} and for valuable discussions. The authors also thank Simeon Bird, George Efstathiou, Andreu Font-Ribera, Joe Hennawi, Chris Pedersen, Andrew Pontzen, Uros Seljak, Licia Verde, Risa Wechsler, Matteo Viel and the members of the \textit{OKC axions} journal club for valuable discussions. KKR thanks the organizers of the ``Understanding Cosmological Observations'' workshop in Benasque, Spain in 2019, where part of this work was performed. HVP acknowledges the hospitality of the Aspen Center for Physics, which is supported by National Science Foundation grant PHY-1607611. KKR was supported by the Science Research Council (VR) of Sweden. HVP was supported by the Science and Technology Facilities Council (STFC) Consolidated Grant number ST/R000476/1 and the research project grant ``Fundamental Physics from Cosmological Surveys'' funded by the Swedish Research Council (VR) under Dnr 2017-04212. This work was supported in part by the research environment grant ``Detecting Axion Dark Matter In The Sky And In The Lab (AxionDM)'' funded by the Swedish Research Council (VR) under Dnr 2019-02337. This work used computing facilities provided by the UCL Cosmoparticle Initiative; and we thank the HPC systems manager Edd Edmondson for his indefatigable support. This work used computing equipment funded by the Research Capital Investment Fund (RCIF) provided by UK Research and Innovation (UKRI), and partially funded by the UCL Cosmoparticle Initiative.
\end{acknowledgments}

\appendix
\section{\label{sec:appendix_convergence}Numerical convergence}

In this analysis, we use a simulation mass resolution and box volume that at least matches (and often improves upon) the numerics of previous Lyman-alpha forest studies of the small-scale matter power spectrum \citep[\(k_\mathrm{f} \sim 10^{-1}\,\mathrm{s}\,\mathrm{km}^{-1}\); \eg][]{2019ApJ...872..101B}; here, we test convergence in these properties. The top panel of Fig.~\ref{fig:systematics} shows a test of convergence in the simulated flux power spectrum with respect to the number of simulation particles, while fixing the comoving volume of the simulation box to \((10\,h^{-1}\,\mathrm{Mpc})^3\). This is equivalently a test of convergence with respect to the mass resolution of the simulation. In each case, we rescale mock spectrum optical depths such that each simulation has the same mean transmitted flux in order to remove this parameter dependence. We observe convergence as the number of particles is increased (and hence the mass resolution is improved). We find the correction from having more particles than our nominal value (\(2 \times 512^3\)) never to be greater than \(10 \%\) and always less than the typical uncertainty in flux power spectrum data \citep[\eg][]{2019ApJ...872..101B}. We carry out this test for fiducial model parameters corresponding to a 2 keV warm dark matter particle, since it is known that convergence criteria are more stringent in the presence of suppressed initial conditions \citep[\eg considering the effect of numerical fragmentation;][]{2014MNRAS.439..300L}. We therefore consider this a robust test of numerical convergence.

The second panel from the top of Fig.~\ref{fig:systematics} shows a test of convergence with respect to the comoving volume of the simulation box, while fixing the mass resolution of the simulation. We observe convergence as the box volume is increased. We find the correction from having a larger box than our nominal value \((10\,h^{-1}\,\mathrm{Mpc})^3\) to be \(\sim 10 \%\) or less, which is less than the typical data uncertainty. We find no statistically significant change in the convergence in the observationally-relevant redshift window \(z \sim 4\) --- 5.

\section{\label{sec:appendix_res}Spectral resolution}

The second panel from the bottom of Fig.~\ref{fig:systematics} shows a test of the systematic error in the flux power spectrum arising from a mis-estimation of the spectral resolution. We find that overestimating the resolution by \(20 \%\) \citep[we take the fiducial full-width-at-half-maximum to be \(6\,\mathrm{km}\,\mathrm{s}^{-1}\), \eg][]{2019ApJ...872..101B} increases by \(\sim 10 \%\) the smallest-scale power spectrum mode that we consider. This arises since the smoothing in the transmitted flux that comes from a finite spectral resolution is over-corrected in the flux power spectrum (see Ref.~\cite{2013A&A...559A..85P} for the full model) and too much power is added on small scales. We find the opposite effect when underestimating the spectral resolution. The typical estimated uncertainty on the spectral resolution is \(\sim 10 \%\), which we find would have no statistically significant effect on the flux power spectrum (\(< 5 \%\)) compared to typical data uncertainty \citep[\(\sim 10 \%\) to \(25 \%\); \eg][]{2019ApJ...872..101B}. This would remain true even if the resolution uncertainty was underestimated by a factor of two.

\section{\label{sec:appendix_mean}Mean flux evolution}

\begin{figure}
\includegraphics[width=\columnwidth]{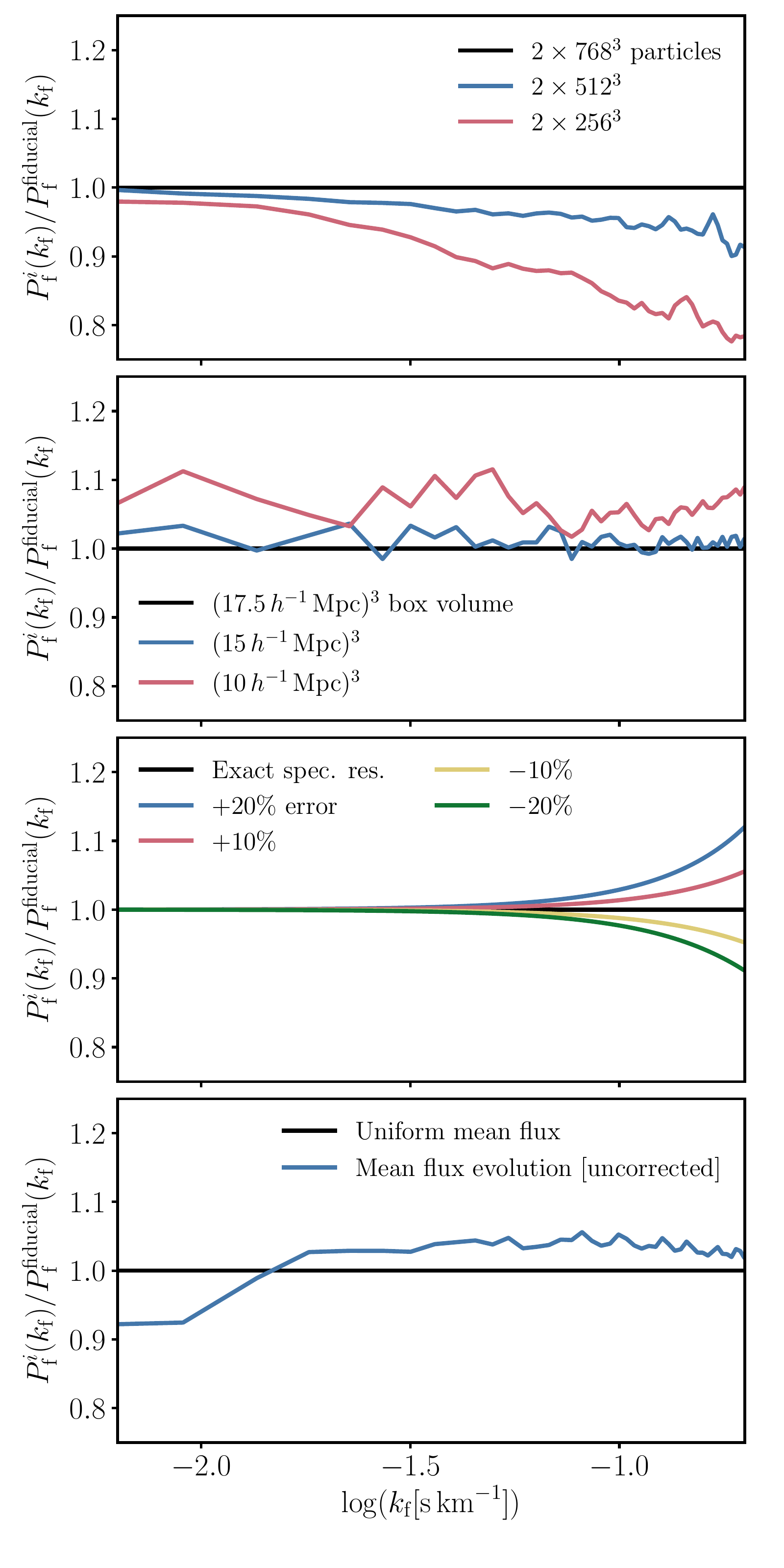}
\caption{\label{fig:systematics}\textit{From top to bottom}: a test of convergence in the simulated flux power spectrum at \(z = 5\) with respect to the number of simulation particles (for fixed comoving box volume); a test of convergence with respect to the comoving volume of the simulation box (for fixed particle mass resolution; the emulator is built with simulations with a comoving volume of \((10\,h^{-1}\,\mathrm{Mpc})^3\) and \(2 \times 512^3\) particles); a test of the systematic error arising from mis-estimating the spectral resolution of the data \citep[up to \(\pm 20 \%\); the uncertainty on the spectral resolution is generally estimated to be \(\sim 10 \%\), \eg][]{2019ApJ...872..101B}; and a test of the systematic error arising from ignoring the redshift evolution of the mean flux in sections of Lyman-alpha forest \citep[the residual error in our analysis will be much less since data are corrected for this effect by the ``rolling mean'' estimator; \eg][]{2019ApJ...872..101B}. For comparison, uncertainty in flux power spectrum data are \(\sim 10 \%\) to \(25 \%\) depending on wavenumber \citep[\eg][]{2019ApJ...872..101B}.}
\end{figure}
The bottom panel of Fig.~\ref{fig:systematics} shows a test of the systematic error in the flux power spectrum at \(z = 5\) arising from ignoring the redshift evolution of the mean transmitted flux in individual sections of Lyman-alpha forest. We model this (without resorting to radiative transfer) by injecting a redshift dependence in the effective optical depth to mock spectra generated from a simulation snapshot at \(z = 5\), using the fiducial model presented in Ref.~\cite{2019ApJ...872..101B}. We then calculate the flux power spectrum from this snapshot ignoring the redshift dependence, \ie following the default procedure of normalizing the flux by the mean in the whole box. We compare this to the flux power spectrum measured from the same snapshot without the injected redshift dependence, but with the optical depths uniformly rescaled to match the global mean flux in the two settings. We find an effect on all scales since the 1D flux power spectrum is a non-linear mapping of the optical depths that we modify. However, as expected, the largest effect is on large scales since we are effectively modifying a long wavelength mode in the optical depth. Nonetheless, the effect on the power spectrum is always \(< 10 \%\) and less than the typical data error \citep[\eg][]{2019ApJ...872..101B}. Further, we expect the residual error in our analysis to be much less than this since the effect is corrected in the data by using a ``rolling mean'' estimator \citep{2019ApJ...872..101B}. Rather than normalizing the transmitted flux by a globally-estimated mean, the mean flux is estimated locally at each pixel using a finite window centered on that point. This corrects for the redshift evolution in the mean flux within sections of Lyman-alpha forest. We therefore expect only a residual error in comparing to simulations at a fixed time-slice. We expect the effect to weaken at lower redshifts. This is because the effective optical depth evolves more slowly at later times and the redshift path length for a given comoving interval also decreases with redshift.

\bibliography{axions}

\end{document}